\documentclass[aps,onecolumn,10pt,pre,showpacs,nofootinbib]{revtex4-1}
\usepackage{graphicx}
\graphicspath{{figures/}{figures/osi/}{figures/finger/}}
\usepackage{overpic}
\usepackage{subfigure}
\usepackage{amsmath}
\usepackage{lineno}
\usepackage[symbol]{footmisc}

\begin{document}


\title{Onset of double-diffusive convection in near-critical gas mixtures}

\author{Zhan-Chao Hu}
\affiliation{Department of Energy and Resources Engineering, College of Engineering, Peking University, Beijing 100871, China.}
\affiliation{Department of Engineering Sciences and Applied Mathematics, Northwestern University, 2145 Sheridan Road, Evanston, Illinois 60208, USA.}
\author{Stephen H.  Davis}
\affiliation{Department of Engineering Sciences and Applied Mathematics, Northwestern University, 2145 Sheridan Road, Evanston, Illinois 60208, USA.}
\author{Xin-Rong Zhang} \email[]{xrzhang@pku.edu.cn}
\affiliation{Department of Energy and Resources Engineering, College of Engineering, Peking University, Beijing 100871, China.}
\newcommand{\RNum}[1]{\uppercase\expandafter{\romannumeral #1\relax}}
\date{\today}

\begin{abstract}
	Near the thermodynamic critical point, the physical properties of binary fluids exhibit large variations in response to small temperature and concentration differences, whose effects on the onset of double-diffusive convection are reported here. The vertical symmetry is broken, irregular penetrative instability occurs, and cat's eye patterns are identified in the fingering regime and oscillatory regime, respectively. A new parameter $\Theta$ is defined which indicates how the variations of physical properties influence flow fields. It is seen through numerical simulations that the Boussinesq approximation with constant physical properties has limited applicability, and that the Boussinesq equations with variable properties and density will describe all features seen. This conclusion is based on comparisons with the fully compressible, variable-property system. 
\end{abstract}


\maketitle

\section{Introduction}\label{sec:intro}
A supercritical fluid is any substance at a temperature and pressure above its liquid-vapor critical point. A near-critical fluid refers to a supercritical fluid with a state slightly above the critical point. It is well-known that close to a critical point, strong anomalies are present in thermodynamic and transport properties of pure fluids \cite{carles2010brief}. Buoyant convection in near-critical pure fluids has attracted much attention \cite{zappoli2016}, in which the (one-component) Rayleigh-B\'{e}nard (RB) problem is a subset. Theoretically, \citet{Carl1999The}, as a validation of the pioneering work of \citet{giterman1970}, pointed out that a modified Rayleigh number which includes the influence of adiabatic temperature gradient (ATG) should be used to predict the onset of convection \cite{Spiegel1971}. Experimental investigations of RB convection \cite{assenheimer1993rayleigh,Kogan1999,meyer2002onset} reveal the development of the convection, the role of ATG and some heat transfer characteristics. A series of numerical simulations has been done \cite{Onuki2001,zappoli2001,ACCARY2009Turbulent,shen2012}, to understand the hydrodynamic behavior of RB convection. The most intriguing phenomenon is that owing to the piston effect, the convection is triggered from both the lower and the upper boundaries when the fluid is heated from below while the top wall keeps its initial temperature.

Like heat transfer, mass transfer in a near-critical fluid also goes with some intriguing phenomena. In our recent work, the onset of natural convection in a near-critical binary gas mixture driven by convection gradient is studied. Due to the ATG, the convection is actually a limiting case of fingering double-diffusive convection (DDC) \cite{hu2018onset}. The criterion for convection onset has been reported there for fingering cases with constant physical properties. This fact inspires us to conduct a comprehensive survey about the onset of DDC in near-critical gases. The highly compressible nature of a near-critical binary gas is anticipated to bring about new features and dynamics. This idea is also prompted by the growing applications of supercritical fluids in chemical extractions, energy technology, and biological synthesis, where near-critical multi-component systems always exist. Therefore, studying DDC in the critical region is fundamental to understand such applications.

DDC is a buoyant instability phenomenon driven by the interaction between two fluid components that diffuse at different rates. The most striking feature of DDC is that convection can be triggered even in a statically stable fluid, namely a fluid that is heavier at bottom. This phenomenon was first discovered by Stern \cite{Stern1960} in an oceanographic application, where temperature and salt concentration are the two competing components. DDC plays an important role in oceanic mixing, and it has also been studied extensively in other applications such as crystal growth, magma dynamics, and planetary interiors \cite{radko2013double}. Depending on the relationship between the diffusivities and gradients of the two components, there are two forms of DDC: fingering and oscillatory. Fingering convection occurs when the slow diffuser is unstably stratified and the fast one is stabilizing. Oscillatory DDC occurs when the fast diffuser is unstably distributed and the slow one is stabilizing. The onset of DDC has been studied through linear stability analysis (LSA) for a horizontal fluid layer subject to vertical temperature and concentration gradients, and stability diagrams have been established for both forms (see for instance \cite{walin1964,nield_1967,baines1969}). 

Depending on the source of the concentration gradient, there are two kinds of problems. In the first kind of problem, the concentration gradient is originated from external factors and is introduced into the model through the boundary conditions. It corresponds to various applications where the supercritical fluids are used as solvents, such as extraction, crystal growth \cite{RASPO20074182} and adsorption \cite{WANNASSI2016203}. Because the Dirichlet boundary condition for concentration is simple, facilitates theoretical studies and helps to extract all phenomena, it has been commonly used by previous studies concerning DDC in other systems, such as \cite{huppert_1976,GOYEAU1996,HAN1991461}. In the second kind of problem, the concentration gradient is established in response to temperature gradient by the cross-diffusion effects (Soret and Dufour effects), where the boundaries are usually considered as impermeable. For near-critical binary fluids, the examples are \cite{NAKANO20074678,LONG2015922}. 

At present, we consider the onset of DDC in a highly compressible near critical binary gas as the first kind of problem. In particular, we investigate the onset of DDC in a near-critical binary gas layer bounded by two horizontal rigid walls and subject to vertical temperature and concentration gradients under Dirichlet boundary conditions. The cross-diffusion effects are omitted. In general, it has been customary to use the Boussinesq approximation (BA) in previous studies of DDC, which brings huge simplifications to the governing equations, and makes it possible to obtain analytical criteria for the onset of convection \cite{turner_1973}. For example, in the pioneering work of \citet{SHTEINBERG1971}, the onset of DDC in near-critical gases is studied based on the Boussinesq equations. However, the applicability of the BA for DDC in near-critical gases is questionable. Here, the applicability of the BA and the influence of variable physical properties on the stability are main concerns, which have not been covered by \cite{SHTEINBERG1971}. Through a numerical LSA, it is shown here that for large enough concentration Rayleigh number, the BA breaks down. The errors can be avoided by modifying the BA by including the variable properties and density.

This paper is organized as follows. The problem under consideration is described in section \ref{sec:pb}. The formulation of LSA is derived in section \ref{sec:flsa}. Sections \ref{sec:fingering} and \ref{osi_Bou} are dedicated to reporting results and discussing the mechanism for the fingering regime and oscillatory regime, respectively. A further discussion on the parameter proposed in the previous two sections is presented in section \ref{sec:fd}. The paper is concluded in section \ref{sec:con}.

\section{Problem description}\label{sec:pb}
\begin{figure}
	\centerline{\includegraphics[width=6cm]{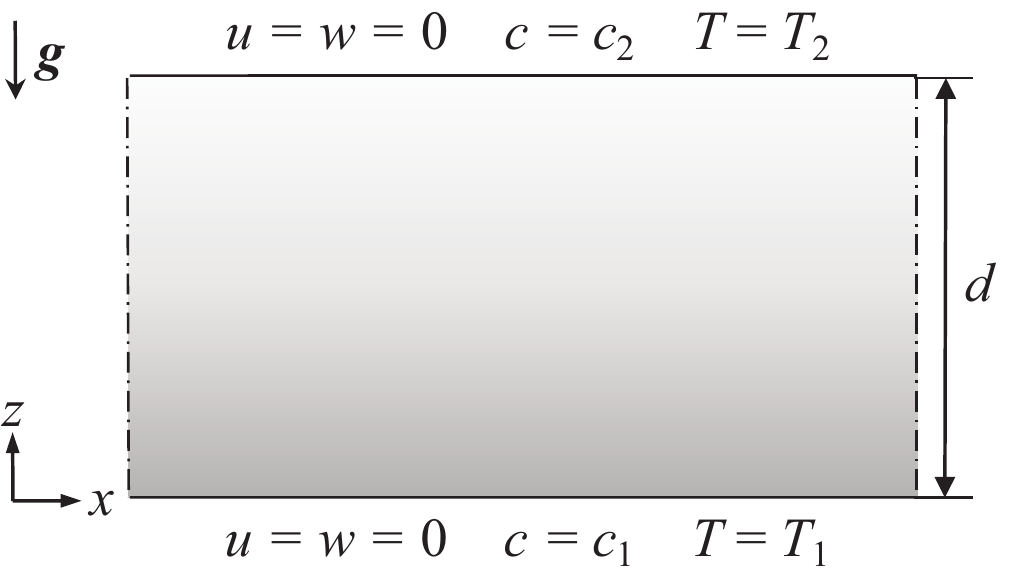}}
	\caption{Physical model for LSA.}
	\label{fig:phy_model}
\end{figure}

Figure \ref{fig:phy_model} shows a two-dimensional horizontal layer with thickness $d$ of a binary gas composed of, say, $\mathrm{CO_2}$ and $\mathrm{C_2H_6}$ close to and above its critical point. The concentration (mass fraction) of $\mathrm{C_2H_6}$ is denoted by $c$. Horizontal and vertical directions, are denoted by $x$ and $z$, respectively. The two horizontal boundaries are rigid, with Dirichlet boundary conditions applied for both temperature and concentration. These are given by
\begin{equation}\label{eq:b_conds}
	{\left. {\begin{array}{*{20}{l}}
				{T = {T_1},\ c=c_1,\ u = w = 0,\ \quad \mbox{on\ }\quad z = 0,}\\[8pt]
				{T = {T_2},\ c=c_2,\ u = w = 0,\ \quad \mbox{on\ }\quad z = d,}
		\end{array}} \right\}}
\end{equation}
where $T$ is the temperature, $u$, $w$ are the horizontal and vertical components of velocity, respectively. The Dirichlet boundary condition for concentration is difficult to attain in experiments. However, in the present theoretical study, it is considered because it is simpler and facilitates to extract all the phenomena.

In the critical region, thermodynamic and transport properties of the gases are very sensitive to the variations of temperature and concentration. Define an average reference state of the gas mixture:
\refstepcounter{equation}
$$
T_r = \frac{T_1+T_2}{2},\ c_r= \frac{c_1+c_2}{2}, \ \rho_r=\rho_c|_{c=c_r},
\eqno{(\theequation{\mathit{a},\mathit{b},\mathit{c}})}
$$
where the subscript $r$ denotes reference properties, and $\rho_c$ is the critical density (a function of $c$). Accordingly, the reference reduced temperature is given by
\begin{equation}
	\frac{T_r-T_c|_{c=c_r}}{T_c|_{c=c_r}}=\varepsilon_r\ll1，
\end{equation}
where $T_c$ is the critical temperature (a function of $c$). In this study, $c_r=0.67$ is taken, because the physical properties are available from references (see Appendix \ref{sec:phy_properties} for details). The critical parameters are $T_c|_{c=c_r} = 297.44~ \mathrm{K}, \rho_c|_{c=c_r} = 244.94 ~\mathrm{kg/m^3}$ .The mixture considered in this study satisfies ${\left( {{{\partial \rho }}/{{\partial c}}} \right)_{p,T}} > 0$, where $\rho$ is the density of the mixture and $ p $ the pressure. The base state density $\bar\rho$ and reference pressure $p_{r}$ satisfy the condition of static equilibrium with the constraint of a mean density equal to $\rho_r$, namely
\refstepcounter{equation}
$$
-\frac{{{\rm d} p_{r}}}{{{\rm d}z}} = \bar\rho g,~
\int_0^d\bar\rho \;{\rm d}z= {\rho _r}d,
\eqno{(\theequation{\mathit{a},\mathit{b}})} \label{con_rho}
$$
with $g=9.81~ \rm{m/s^2}$, the gravitational acceleration. 

The governing equations for binary gases developed by Landau and Lifshitz \cite{landau1987fluid} together with a linearized equation of state are adopted and modified to describe a Newtonian, compressible, viscous and heat-conducting near-critical binary gas:
\begin{eqnarray}\label{eq:ge}
	&&\frac{{\partial \rho }}{{\partial t}} + \nabla  \cdot (\rho {\mathbf{u}}) = 0,\nonumber \\
	&&\rho \left[ {\frac{{\partial {\mathbf{u}}}}{{\partial t}} + ({\mathbf{u}} \cdot \nabla ){\mathbf{u}}} \right] =  - \nabla p + \nabla  \cdot (\eta \nabla {\mathbf{u}}) + \nabla  \cdot [\eta {(\nabla {\mathbf{u}})^\mathrm{T}}] - \frac{2}{3}\nabla (\eta \nabla  \cdot {\mathbf{u}}) + (\rho  - \bar \rho ){\mathbf{g}}, \nonumber \\
	&&\rho {c_p}\left( {\frac{{\partial T}}{{\partial t}} + {\mathbf{u}} \cdot \nabla T} \right) = \nabla  \cdot (\lambda \nabla T) + \frac{{{\rho _r}\beta T}}{\rho }\left( {\frac{{{\rm{D}}p}}{{{\rm{D}}t}} - \bar \rho gw} \right)+\Phi,\\
	&&\rho \left( {\frac{{\partial c}}{{\partial t}} + {\mathbf{u}} \cdot \nabla c} \right) = \nabla \cdot(\rho D\nabla c), \nonumber\\
	&&\rho  = \bar \rho  + {\rho _r}\left[ {\alpha p - \beta (T - \bar T) + \kappa (c - \bar c)} \right], \nonumber
\end{eqnarray}
where $t$ is the time; velocity ${\mathbf{u}}=(u,w)$; $\eta$ and ${\mathbf{g}}(0,-g)$ are, respectively, the dynamic viscosity and the gravitational acceleration vector; $D$ is the diffusion coefficient; $ c_p $ and $\lambda$ are the specific heat at constant pressure and the thermal conductivity, respectively; $\Phi  = \eta [\nabla {\mathbf{u}} + {(\nabla {\mathbf{u}})^{{T}}}]:\nabla {\mathbf{u}} - {2}/{3}\eta {(\nabla  \cdot {\mathbf{u}})^2}$ is the viscous heating function; ${\alpha} = \rho_r^{-1}{\left( {\partial \rho /\partial p} \right)_{T,c}}$ is the isothermal compressibility; ${\beta} =  - \rho_r^{-1}{\left( {\partial \rho /\partial T} \right)_{p,c}}$ is the isobaric thermal expansion coefficient; ${\kappa} =  \rho_r^{-1} {\left( {\partial \rho /\partial c} \right)_{p,T}}$ is the concentration contraction coefficient. In the derivations of the above equations, the bulk viscosity is neglected \cite{zappoli2001}. In addition, a 
decomposition of total pressure
\begin{equation}\label{eq:dec_p}
p_{total}=p_{r}+p,
\end{equation}
along with Eq. (\ref{con_rho}$a$) is employed in the derivations of energy and concentration conservation equations.

Based on previous studies, the ATG plays an important role in the RB convection for near-critical pure gases \cite{SHTEINBERG1971,Carl1999The}. As a result, a modified thermal Rayleigh number is used, where the role of temperature gradient is replaced by its departure from the adiabatic one \cite{Spiegel1971,Carl1999The}. These facts suggest defining the following non-dimensional quantities (denoted by the primes):
\begin{eqnarray}\label{eq:def_non}
	&&(x,z) = (x',z') \cdot d, ~~t = t' \cdot \frac{{{d^2}}}{{{{D_{T,r}}}}},~~\mathrm{u} (u,w) = \mathrm{u'}(u',w')\cdot \frac{{D_{T,r}}}{{{d}}},\nonumber\\
	&&T = T'\cdot{(\nabla_{T}-\nabla_{ad,r})}d + {T_r}, ~~c = c' \cdot \nabla_{c}d+c_r, \nonumber\\
	&&p = p' \cdot \frac{{{\eta_r {{D_{T,r}}}}}}{{{d^2}}}, ~~(\rho, \bar \rho) = (\rho', \bar {\rho'})  \cdot {\rho _r}, \\
	&& \eta = \eta'\cdot\eta_{r} ,~~ \lambda = \lambda'\cdot \lambda_r, ~~ D = D'\cdot{D_{r}}, \nonumber\\
	&& c_p = c_p'\cdot c_{p,r},~~\alpha = \alpha'\cdot\alpha_{r}, ~~\beta = \beta'\cdot\beta_r, ~~\kappa = \kappa'\cdot\kappa_r, \nonumber
\end{eqnarray}
where ${D_{T,r}}=\lambda_r/(\rho_r {c_{p,r}})$, $\nabla_{T}=({T_1-T_2})/{d}$, $\nabla_{c}=({c_1-c_2})/{d}$, $\nabla_{ad,r} = T_r\beta_{r}g/c_{p,r}$ (the reference ATG). By substituting the system (\ref{eq:def_non}) into the system (\ref{eq:ge}), the non-dimensional governing equations are obtained as 
\begin{eqnarray}\label{eq:non-full}
&&\frac{{\partial \rho' }}{{\partial t'}} + \nabla  \cdot (\rho' {\mathbf{u'}}) = 0,\nonumber\\
&&\frac{{\rho '}}{{\rm Pr }}\left[ {\frac{{\partial {\mathbf{u'}}}}{{\partial t'}} + ({\mathbf{u'}} \cdot \nabla ){\mathbf{u'}}} \right] =  - \nabla p' + \nabla  \cdot \left( {\eta'\nabla \mathbf{u'}} \right) + \nabla  \cdot \left[ {\eta'{{(\nabla \mathbf{u'})}^{\rm{T}}}} \right] - \frac{2}{3}\nabla \left( {\eta'\nabla  \cdot \mathbf{u'}} \right) \nonumber\\
&&\qquad\qquad\qquad\;\quad\qquad\qquad+ \left[ {\chi\alpha' p' - {\rm Ra_T}\beta' (T' - \bar T') + {\rm Ra_S}\kappa' (c' - \bar c')} \right]{\mathbf{k}},\nonumber \\
&& \frac{{\partial T'}}{{\partial t'}} + {\mathbf{u}}' \cdot \nabla T' = \frac{1}{{\rho '{c_p}'}}\left[ {\nabla (\lambda '\nabla T') + \frac{{{{Ra}_{ad}}}}{{\rm Ra_T}}\frac{{\beta '}}{{\rho '}}\left( {{ \Pi}\frac{{Dp'}}{{Dt'}} - \bar \rho 'w'} \right)} \right] +\Phi', \\
&& {\frac{{\partial c'}}{{\partial t'}} + \mathbf{u'} \cdot \nabla c'} = \frac{\tau}{\rho'}\nabla \cdot(\rho' D'\nabla c'),\nonumber\\
&&\rho'  = \bar \rho'  +  \Pi\left[ {\chi\alpha' p'- {\rm Ra_T}\beta' (T' - \bar T') + {\rm Ra_S}\kappa' (c' - \bar c')} \right].\nonumber
\end{eqnarray}
Expressions and the physical meanings of non-dimensional numbers in above equations are summarized in Table \ref{Tab:non-numbers}. Apart from $\rm Ra_T$, $\rm Ra_S$, $\rm Pr $ and $\tau$ as in DDC for Boussinesq fluids, three new non-dimensional numbers are identified. The adiabatic Rayleigh number is the Rayleigh number in terms of ATG. It naturally arises from the modification of the thermal Rayleigh number, and $\rm Ra_T+Ra_{ad}$ yields the classical thermal Rayleigh number. The compressibility factor, denoted by $\chi$, represents the non-dimensional density stratification induced by gravity. The pressure ratio, denoted by $ \Pi$, stands for the relative importance of dynamic pressure and hydrostatic pressure. 

Note that in Eq. (\ref{eq:non-full}), the variable adiabatic Rayleigh number ${Ra}_{ad}$ (differentiated by italic) is present, where the reference ATG, denoted by $\nabla_{ad,r}$, is replaced by the variable one:
\begin{equation}
\nabla_{ad}=\frac{T\beta_rg}{c_{p,r}}.
\end{equation}
\begin{table}
	\caption{\label{Tab:non-numbers} A summary of the non-dimensional numbers.}
	\begin{ruledtabular}
		\begin{tabular}{ccl}
			Notation& Expressions&Name\\
			\hline
			$\rm Ra_T$ & $\dfrac{{{\rho _r}g\beta_r {(\nabla_{T}-\nabla_{ad,r})}{d^4}}}{{\eta_r D_{T,r}}}$ & Thermal Rayleigh number. \\
			$\rm Ra_S$ & $\dfrac{{{\rho _r}g\kappa_r {\nabla_{c}}{d^4}}}{{\eta_r D_{T,r}}}$ & Concentration Rayleigh number. \\
			$\rm Ra_{ad}$ & $\dfrac{{{\rho _r}g\beta_r {\nabla_{ad,r}}{d^4}}}{{\eta_r D_{T,r}}}$ & Adiabatic Rayleigh number. \\
			$\tau$& $D_r/D_{T,r}$& Diffusivity ratio.\\
			$\rm Pr $& $\eta_r/(\rho_rD_{T,r})$& Prandtl number.\\
			$\chi$&$\alpha_r {\rho _r}gd$& Compressibility factor.\\
			$\Pi$ & $ \dfrac{{\eta_r D_{T,r}}}{{\rho_rg{d^3}}}$ & Pressure ratio.\\
		\end{tabular}
	\end{ruledtabular}
\end{table}

For the sake of simplicity, the prime symbol ($'$) for any non-dimensional variable is omitted hereafter.

\section{Formulation of linear stability theory}\label{sec:flsa}
\begin{figure}
	\centering
	\includegraphics[width=12cm]{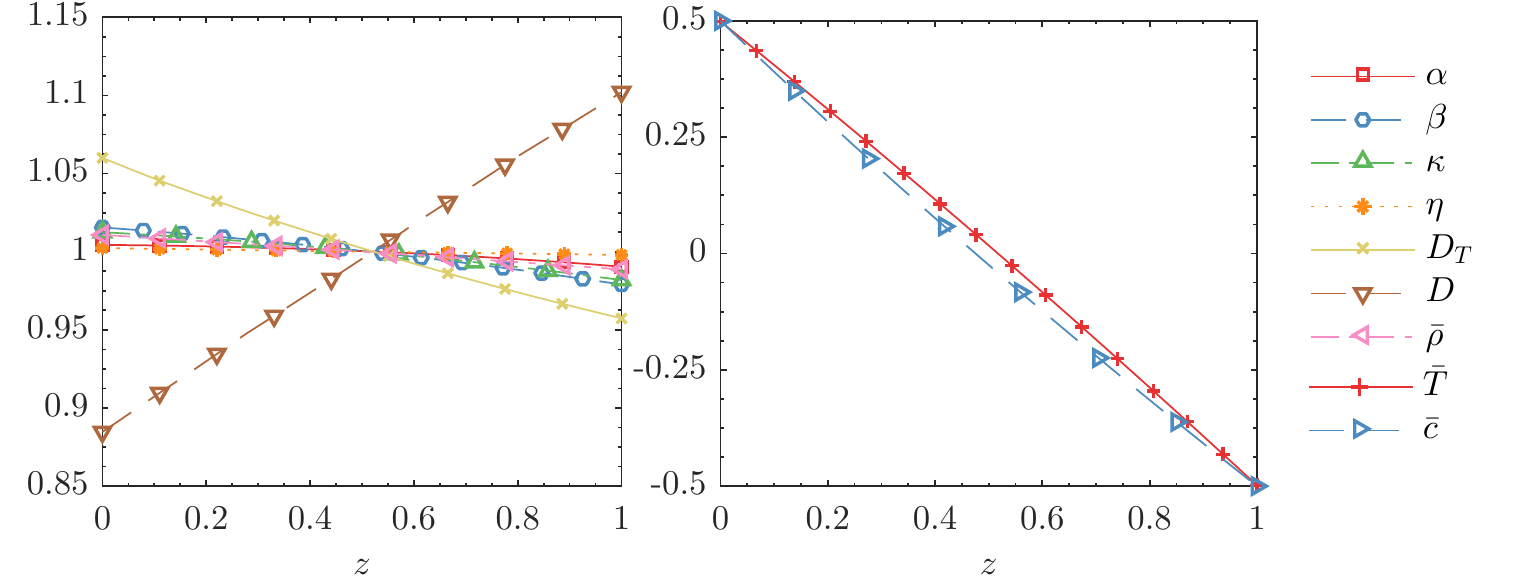}
	\caption{Physical properties and state variables against $z$ for ${\rm Ra_S} = -1\times10^7$ and  ${\rm Ra_T}=-1.03\times10^8$.}
	\label{fig:variable_properties_fin}
\end{figure}

\begin{figure}
	\centering
	\includegraphics[width=12cm]{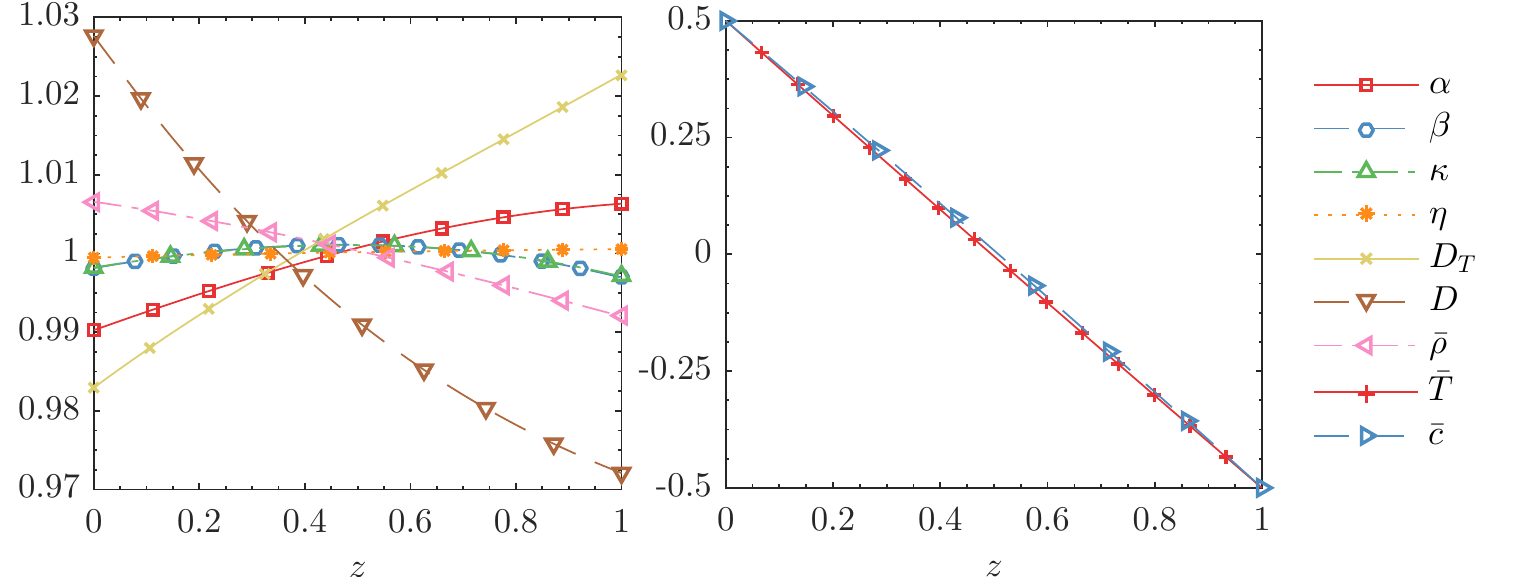}
	\caption{Physical properties and state variables against $z$ for ${\rm Ra_S} = 3.67\times10^8$ and ${\rm Ra_T}=3.06\times10^8$.}
	\label{fig:variable_properties_osi}
\end{figure}

Consider a stationary base state, namely
\begin{equation}\label{eq:base_u}
\mathrm{\bar u}(\bar u,\bar w)= (0,0),
\end{equation}
where the bar symbol denotes initial values. Eq. (\ref{eq:dec_p}) gives 
\begin{equation}
\bar p=0.
\end{equation}
Because physical quantities are nonlinear functions of $T$, $c$ and $\rho$, the classical assumption of constant concentration and temperature gradients no longer holds. According to Eqs. (\ref{con_rho}), (\ref{eq:non-full}) and (\ref{eq:base_u}), the non-dimensional initial fields for $T$, $c$ and $\rho$ obey
\begin{eqnarray}\label{eq:base_Tc}
	\chi \alpha \bar \rho + \frac{{{\rm{d}}\bar \rho }}{{{\rm{d}}z}} + { \Pi}\left( {{\rm Ra_T}\beta \frac{{{\rm{d}}\bar T}}{{{\rm{d}}z}} - {{\rm Ra_S}}\kappa \frac{{{\rm{d}}\bar c}}{{{\rm{d}}z}}} \right) &=& 0,\nonumber \\
	\lambda \frac{{{{\rm{d}}^2}\bar T}}{{{\rm{d}}z{^2}}} + \frac{{{\rm{d}}\lambda }}{{{\rm{d}}z}}\frac{{{\rm{d}}\bar T}}{{{\rm{d}}z}} &=& 0,\\
	\bar \rho D\frac{{{{\rm{d}}^2}\bar c}}{{{\rm{d}}z{^2}}} + \frac{{{\rm{d}}\bar \rho D}}{{{\rm{d}}z}}\frac{{{\rm{d}}\bar c}}{{{\rm{d}}z}} &=& 0, \nonumber \\
	\int_0^1 {\bar \rho } \,{\rm{d}}z &=& 1.\nonumber
\end{eqnarray}
The base state is closely coupled with the variations in physical properties. The above equations are solved by finite differences to obtain $\bar T(z)$, $\bar c(z)$ and $\bar \rho(z)$. The numerical method is presented in our recent work \cite{hu2018onset}. The modeling of physical properties is elaborated in Appendix \ref{sec:phy_properties}, together with a brief review about the critical behavior of physical properties.  Figures \ref{fig:variable_properties_fin} and \ref{fig:variable_properties_osi} show the profiles of state variables and physical properties for two typical cases: ${\rm Ra_S} = -1\times10^7$ \& ${\rm Ra_T}=-1.03\times10^8$ and  ${\rm Ra_S} = 3.67\times10^8$ \& ${\rm Ra_T}=3.06\times10^8$, respectively. All physical properties and state variables vary nonlinearly along $z$. Besides, it is demonstrated in these two figures that the variations in $D$ exceed those of other properties.

Then subject the base state to small disturbances (denoted by tildes):
\begin{equation}\label{eq:pert}
{\mathrm{u}} = {\mathrm{\bar u}} + {\mathrm{\tilde u}},~~T=\bar T+\tilde T,~~c=\bar c+\tilde c,~~p=\bar{p}+\tilde p,~~\rho=\bar \rho+\tilde\rho.
\end{equation}
Substituting Eq. (\ref{eq:pert}) into Eq.  (\ref{eq:non-full}), omitting high-order nonlinear terms, and letting
\begin{equation}\label{eq:normal_modes}
[\tilde u,\tilde w,\tilde p,\tilde T,\tilde c,\tilde \rho] = [\hat u(z),\hat w(z),\hat p(z),\hat T(z),\hat c(z),\hat \rho(z)]{\rm exp}(ikx+\sigma t),
\end{equation}
one arrives at
\begin{eqnarray}\label{eq:LSA_var}
	\eta\left( {{{\mathcal{D}}^2} - \frac{4}{3}{k^2}} \right)\hat u + {\eta _z}{\mathcal{D}}\hat u + \frac{1}{3}ik\eta {\mathcal{D}}\hat w && \nonumber\\
	+ik{\eta_z}\hat w - ik\hat p &=& \sigma \left( {\frac{1}{{\rm Pr  }}\bar \rho \hat u} \right),\nonumber\\
	\frac{1}{3}ik\eta {\mathcal{D}}\hat u - \frac{2}{3}ik{\eta _z}\hat u + \eta \left( {\frac{4}{3}{{\mathcal{D}}^2} - {k^2}} \right)\hat w +  \frac{4}{3}{\eta _z}{\mathcal{D}}\hat w&& \nonumber\\
	+ {\rm Ra_T}\beta \hat T - {{\rm Ra_S}}\kappa \hat c - ({\mathcal{D}} + \alpha \chi )\hat p &=& \sigma \left( {\frac{1}{{\rm Pr  }}\bar \rho \hat w} \right), \nonumber\\
	- \left( {{{\bar T}_z} + \frac{{{\bar{{Ra}}_{ad}}}}{{\rm Ra_T}}\frac{\beta }{{\bar \rho {c_p}}}} \right)\hat w + \frac{\lambda }{{\bar \rho {c_p}}}({{\mathcal{D}}^{\rm{2}}} - {k^2})\hat T + \frac{{{\lambda _z}}}{{\bar \rho {c_p}}}{\mathcal{D}}\hat T &=& \sigma \left( {\hat T - \frac{\beta }{{{{\bar \rho }^2}{c_p}}}\frac{{{ \Pi}{\bar{{Ra}}_{ad}}}}{{\rm Ra_T}}\hat p} \right),\\
	- {{\bar c}_z}\hat w + \tau D({{\mathcal{D}}^{\rm{2}}} - {k^2})\hat c + \frac{\tau }{{\bar \rho }}{(\bar \rho D)_z}{\mathcal{D}}\hat c &=& \sigma \hat c,\nonumber\\
	- ik\bar \rho \hat u -(\bar \rho_z  + \bar \rho {\mathcal{D}} ) \hat w&=& \sigma \left[ {{ \Pi}(\chi \alpha \hat p - {\rm Ra_T}\beta \hat T + {{\rm Ra_S}}\kappa \hat c)} \right],\quad\quad \nonumber
\end{eqnarray}
where $\sigma$ is the growth rate, $k$ is the horizontal wave number, the operator $\mathcal{D}={\rm d}/{\rm d }z$, and the subscript $z$ denotes a derivative with respect to $z$. The corresponding boundary conditions are
\begin{equation}\label{eq:b_full}
\hat u = \hat w = \hat T = \hat c = 0,  \ \quad \mbox{on\ }\quad z = 0 ~\&~ 1.
\end{equation}
Due to its complexity, the above system of equations is solved numerically to obtain the disturbance profiles and $\sigma$ for a given $k$. The numerical method is detailed in our previous work \cite{hu2018onset}.


To study the applicability of the BA and the influences of variable properties, four reduced cases of system (\ref{eq:LSA_var}) are presented in Appendix \ref{appB} and summarized briefly in Table \ref{tab:sum_cases}, 
\begin{table}
	\begin{ruledtabular}
		\begin{tabular}{cclc}
			Case  & Equation & Implication & Abbreviation \\[3pt]
			\hline
			\RNum{1}  & Eq. (\ref{eq:case1}) & full equations with constant properties & FC \\
			\RNum{2}  & Eq. (\ref{eq:case2})& Boussinesq equations with constant properties & BC \\
			\RNum{3}  & Eq. (\ref{eq:case3})& modified Boussinesq approximation with constant properties & BC' \\
			\RNum{4}  & Eq. (\ref{eq:LSA_var}) & full equations with variable properties & FV \\
			\RNum{5}  & Eq. (\ref{eq:case5}) & modified Boussinesq approximation with variable properties & BV'\\
		\end{tabular}
		\caption{A summary of cases considered in this study.}
		\label{tab:sum_cases}
	\end{ruledtabular}
\end{table}
where FC represents the model of full equations with constant properties, BC the Boussinesq equations with constant properties, BC' the modified Boussinesq equations with constant properties, FV the full equations with variable properties, and BV' the modified Boussinesq equations with variable properties. These abbreviations are used to refer to them in later sections.

The current problem is governed by five parameters: $\rm Ra_T$, $\rm Ra_S$, $k$, $\varepsilon_r$ and $d$. 
Since LSA is performed, this study focuses on neutral stability.  For a given $\rm Ra_S$, $\varepsilon_r$ and $d$, the neutral stability is defined by the minimum $\rm Ra_T$ satisfying  ${\rm Re}(\sigma)=0$ for the most unstable mode of a critical wave number, and ${\rm Re}(\sigma)<0$ for all other wave numbers. Then $\rm Ra_S$ is changed and above procedures are repeated to depict the neutral stability curve. In order to save CPU time, the golden-section search technique is employed in the algorithm. 
In later sections, for the sake of simplicity, $\rm Ra_T$ and $k$ particularly represent the critical $\rm Ra_T$ and critical $k$, respectively. And $\sigma$ denotes the complex growth rate of the most unstable mode.

A reference reduced temperature $\varepsilon_r = 0.001$ and a height $d=10$ mm are set throughout this study, which yields $\rm Ra_{ad}=2.991\times10^5$, $\tau= 0.127$, $\rm Pr = 4.066$, $\chi=7.405\times10^{-5}$ and $ \Pi=2.378\times10^{-10}$. Besides, due to $\tau<1$, oscillatory instability occurs when ${\rm Ra_S}>0$ and ${\rm Ra_T}>0$, while fingering instability occurs when ${\rm Ra_S}<0$ and ${\rm Ra_T}<0$. The results and discussions are presented in the following sections. We begin with the fingering regime.

\section{Fingering regime}\label{sec:fingering}
\subsection{The applicability of the Boussinesq approximation}\label{sec:fin_bou}
\begin{figure}
	\centering
	\includegraphics[width=13cm]{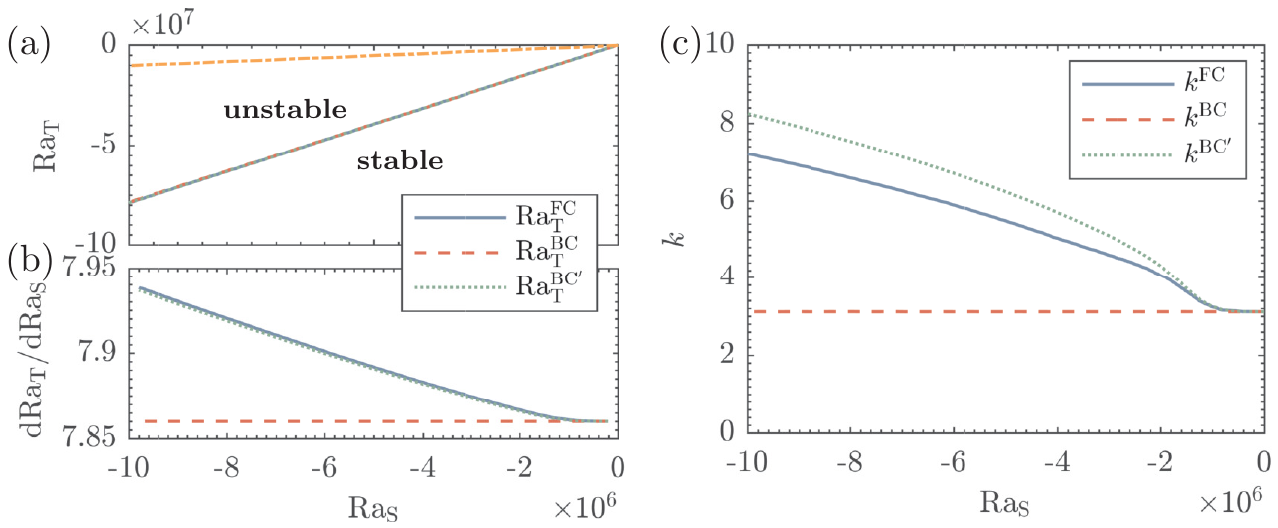}
	\caption{Comparisons among FC, BC and BC' in the fingering regime. (a): Neutral curves satisfying ${\rm Re}(\sigma)=0$, (b): slopes of the neutral curves, (c): critical wave numbers, with solid lines for case FC, dashed lines for case BC and dotted lines for case BC'. The dash-dot line shown in figure (a) represents the boundary for statically stable, below which the gas is heavy at bottom.}
	\label{fig:comp_finger_123}
\end{figure}

The purpose of this section is to investigate the applicability of the BA. In this regard, FC, BC and BC' are compared and discussed. All physical properties are constant. On the neutral curve of fingering regime,  $\sigma=0$. These three cases are compared in terms of $\rm Ra_T$ and $k$ in Fig. \ref{fig:comp_finger_123}. The range of $\rm Ra_S$ is $[-1\times 10^7,0]$. In fact, part of this problem is addressed in our previous work \cite{hu2018onset}, where the applicability of the criterion obtained from the BA for onset of convection is examined for several values of $\rm Ra_S$. This section complements our previous work and helps to draw a whole picture under the current configuration.

\begin{figure}
	\centering
	\includegraphics[width=6cm]{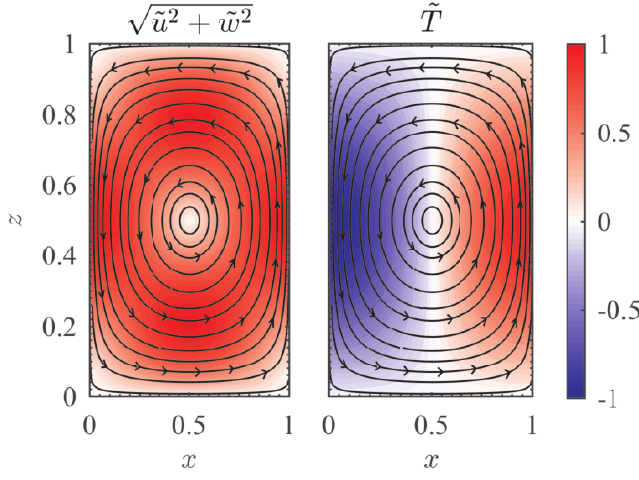}
	\caption{Contours of magnitude of velocity perturbation $|\mathbf{\tilde u}|=\sqrt{{\tilde u}^2+{\tilde w}^2}$ and temperature perturbation $\tilde T$ at the onset of fingering convection calculated from BC, in which $|\mathbf{\tilde u}|$ is scaled to $[0,1]$, and $\tilde T$ to $[-1,1]$. Flow structure is shown by streamlines, with arrows denoting the direction of flow. The range of $x$ presented is $[0,1]$. }
	\label{fig:fields_fin_2}
\end{figure}
First regard the results obtained from BC. It is evident that in BC, after the BA is applied, the neutral curve is a straight line (a constant slope in Fig. \ref{fig:comp_finger_123}(b)) and $k$ is a constant value equaling 3.116, which are in good agreement with classical results \cite{Reid1958,nield_1967}. Figure \ref{fig:fields_fin_2} shows the profiles of temperature perturbation ${\tilde T}$ and magnitude of velocity perturbation $|\mathbf{\tilde u}|$, along with the flow structure at the onset of fingering instability obtained from BC. For the sake of clarity, $\mathbf{\tilde u}$ is scaled to $[0,1]$ and ${\tilde T}$ to $[-1,1]$. The physical space in the $z$-direction is occupied by a single vortex. The temperature perturbation is controlled by vertically convective transport, with a profile proportional to $\tilde w$.

Comparing FC to BC of Fig. \ref{fig:comp_finger_123}, it is important to note that differences are observed when $|{\rm Ra_S}| > 1\times 10^6$. The actual neutral curve in FC is no longer a straight line but a downward concave, whose slope increases monotonically with $|{\rm Ra_S}|$ (Fig. \ref{fig:comp_finger_123}(b)). On the other hand, the critical wave number at neutral stability is also no longer a constant, but increases with $|{\rm Ra_S}|$. 

\begin{figure}
	\centering
	\includegraphics[width=13cm]{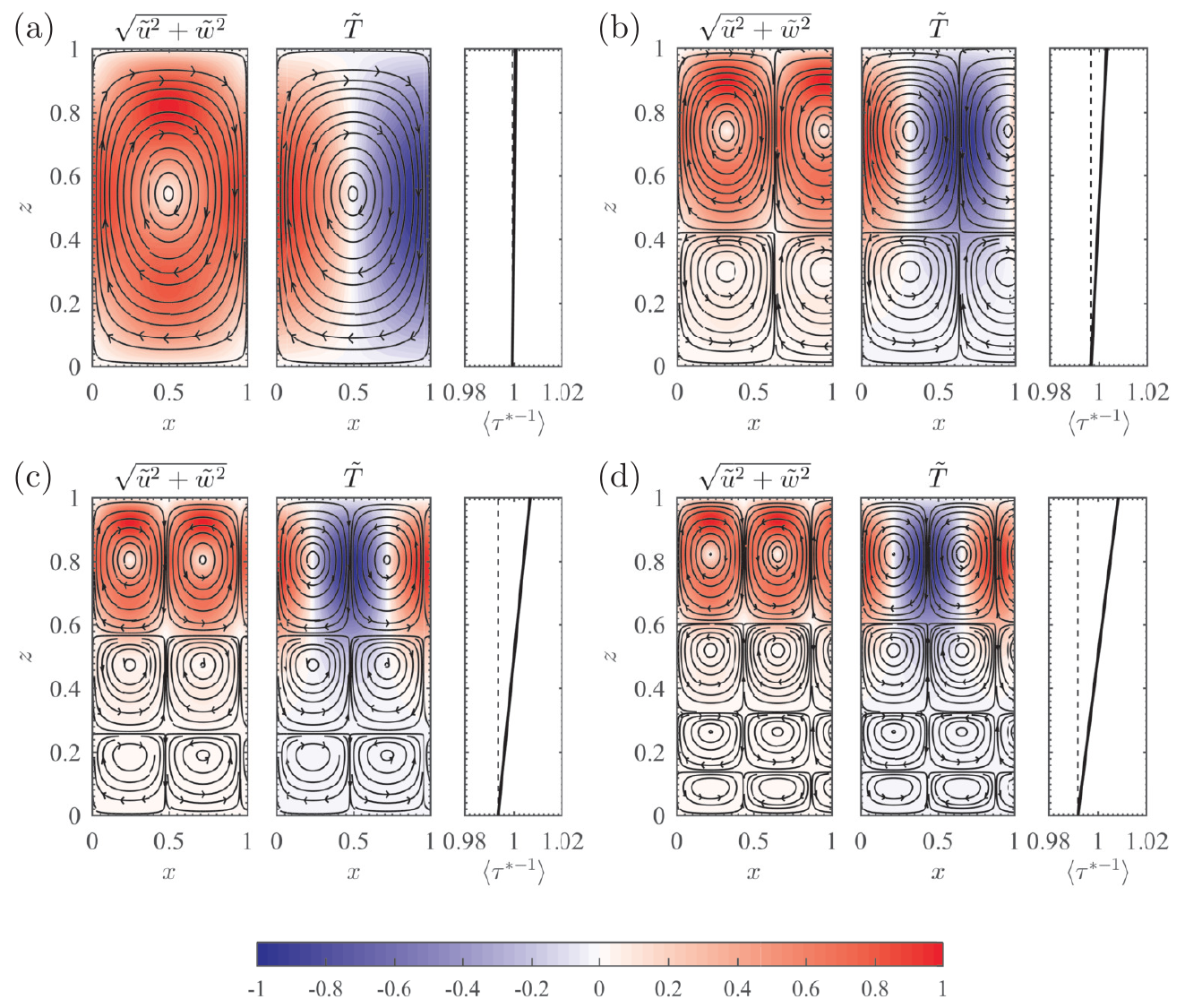}
	\caption{See the caption of Fig. \ref{fig:fields_fin_2} for the description of contours and streamlines. The presented results are obtained from FC with (a) ${\rm Ra_S} = -8.16\times10^5$, (b) ${\rm Ra_S} = -3.88\times10^6$, (c) ${\rm Ra_S} = -7.96\times10^6$ and (d) ${\rm Ra_S} = -1\times10^7$. $\tau^{*}$ is the actual thermal diffusivity given by Eq. (\ref{eq:tau_sat}). $\langle\tau^{*-1}\rangle = \tau^{*-1}/\tau^{*-1}_{m}$, with $\tau^{*-1}_{m}$ denoting its mean value. To make the vertical non-uniformity of $\langle\tau^{*-1}\rangle$ clear, additional dashed lines representing the minimum value of  $\langle\tau^{*-1}\rangle$ are plotted.}
	\label{fig:fields_fin_1}
\end{figure}
To visualize the differences described above, Fig. \ref{fig:fields_fin_1} shows the evolution of fields for different $\rm Ra_S$ along the neutral curve in FC. In Fig. \ref{fig:fields_fin_1}(a) with ${\rm Ra_S} = -8.16\times10^5$, no obvious deviations from Fig. \ref{fig:fields_fin_2} are observed, implying the BA is still reliable. However, as $\rm Ra_S$ is further increased, a layering structure is observed. In the $z$-direction, the previous single vortex is replaced by several organized vortices; the physical space is layered. Every layer is filled with vortices with the same size and intensity, and vortices in two adjacent layers rotate in opposite directions. The convective transport in the top layer is strongest, and both the size and intensity of the vortices decrease downward. Besides, the layering is enhanced as $|{\rm Ra_S}|$ is increased. As presented by Figs. \ref{fig:fields_fin_1}(b) - \ref{fig:fields_fin_1}(d), when $\rm Ra_S$ varies from $-3.88\times10^6$ to $-1\times10^7$, the number of layers increases from 2 to 4.

In fact, the layering introduced above is an example of `penetrative instability', which happens when an unstable region is bounded by stable regions \cite{Lionel1967}. Here, the top layer defines the unstable region, and the convective motion in it penetrates into the neighboring stable region through the shearing force. 

Our previous arguments in \cite{hu2018onset} suggest that it is the omitting of base-state density stratification that leads to the failure of the BA. In order to confirm this reasoning, consider a modification to the BA by including the base-state density stratification. Final equations are denoted as BC' given by Eq. (\ref{eq:case3}). The corresponding results have been shown in Fig. \ref{fig:comp_finger_123} (dotted curves). After the base-state density stratification is included, the BA is greatly improved. On the one hand, the bending of the neutral curve is reproduced, and its slope is predicted accurately (Fig. \ref{fig:comp_finger_123}(b)). On the other hand, the monotonic increase of wave number is also obtained, although there are still obvious differences compared to FC.

\begin{figure}
	\centering
	\includegraphics[width=6cm]{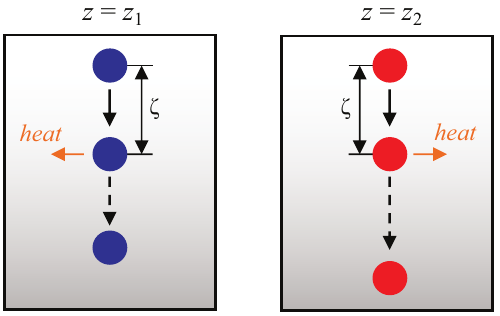}
	\caption{Illustration of the effect of base-state density stratification in the fingering regime.}
	\label{fig:ill_fin_rho}
\end{figure}

To understand the mechanism, the actual dimensional thermal diffusivity is a key property \cite{hu2018onset},
\begin{equation}\label{eq:real_DT}
D_T^\ast(z)=\frac{1}{\bar\rho}\frac{\lambda_r}{ \rho_rc_{p,r}}=\frac{1}{\bar\rho}D_{T,r}. 
\end{equation}
Figure \ref{fig:comp_finger_123}(a) suggests that convection is triggered in a statically stable state, namely, $\bar\rho_z<0$. Since $D_{T,r}$ is a constant, $D_T^\ast$ increases upward due to the reciprocal of $\bar\rho$. A thought experiment presented in Fig. \ref{fig:ill_fin_rho} helps elaborate the mechanism. Consider two parcels of fluid initially at  $z=z_1$ and $z_2$, with $z_1<z_2$, the distribution of $D_T^\ast$ gives $D_{T1}^\ast< D_{T2}^\ast$. Perturb both of them downward by an infinitesimal distance $\zeta$. Then, at the new positions, both of them are hotter and have a higher concentrations than their surroundings. Due to $\tau\ll1$, they adjust their temperatures quickly through thermal diffusion but their concentrations remain almost unchanged. As a result, they are heavier than their surroundings and will keep dropping. For the upper fluid parcel, because of $D_{T1}^\ast< D_{T2}^\ast$, it loses more heat than the lower one. As a result, it drops faster, as illustrated in Fig. \ref{fig:ill_fin_rho}. The above reasoning implies when $D_{T}\ast$ increases with $z$, the region close to $z=1$ is more capable of amplifying small perturbations, or equivalently, the region close to $z=1$ is neutrally stable, whereas the other region is stable.

Define an actual diffusivity ratio $\tau^\ast$ as \cite{hu2018onset}
\begin{equation}\label{eq:tau_sat}
\tau^\ast(z)=\frac{D_{r}}{D_T^\ast}=\bar\rho\tau. 
\end{equation}
The distributions of $\tau^{\ast-1}$ are plotted against $z$ in Figs. \ref{fig:fields_fin_1}. It is highlighted that $\tau^{\ast-1}$ is a good indicator for the penetrative instability. The unstable region near the top wall is always the region with large $\tau^{\ast-1}$, and the number of layers is proportional to the non-uniformity of $\tau^{\ast-1}$. 

The penetrative instability introduced above happens when all physical properties are constant. As seen later, it is concealed by the effects of variable properties. However, this is still useful. Indeed, a possibility for the breakdown of BA is confirmed and a method to improve it is thus developed. Besides, as a limiting case for the problem studied in the next section (all physical properties will be $z$-dependent), some conclusions in this section will be useful later.

\subsection{The effects of variable properties}\label{sec:vp_fin}
As discussed in section \ref{sec:phy_properties}, physical properties of near-critical gases are very sensitive to the change of state variables. Therefore, inhomogeneities in concentration and temperature should have profound influences on DDC, which is the main topic of this section. To this end, FC, FV and BV' are involved and discussed. FV and BV' are the counterparts of FC and BC' with variable properties, respectively.

\begin{figure}
	\centering
	\includegraphics[width=13cm]{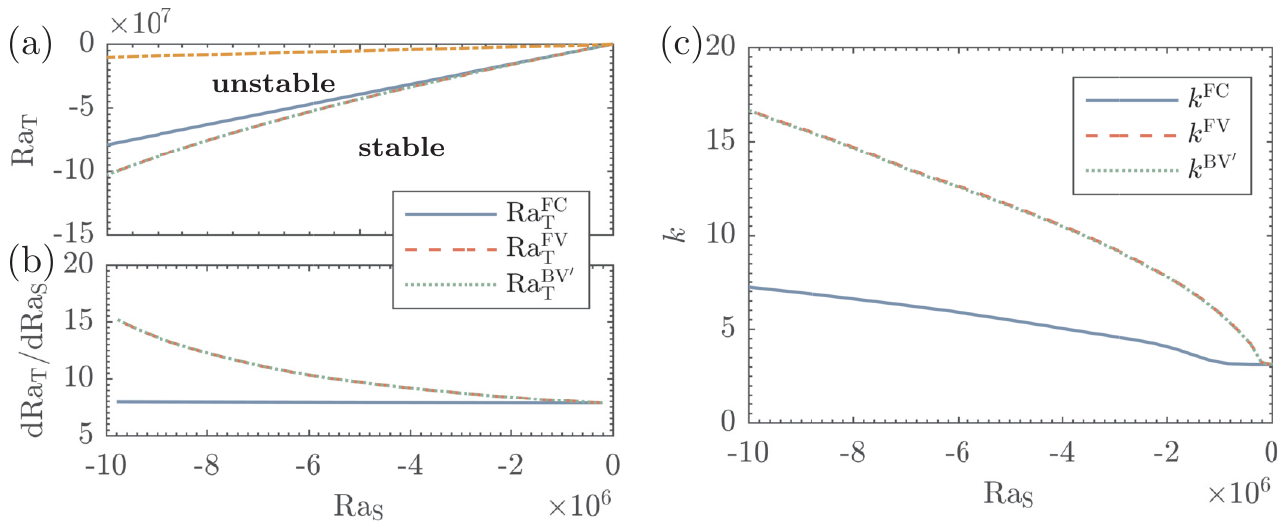}
	\caption{Comparisons among FC, FV and BV' in the fingering regime. (a): Neutral curves satisfying ${\rm Re}(\sigma)=0$, (b): slopes of the neutral curves, (c): critical wave numbers, with solid lines for FC, dashed lines for FV and dotted lines for BV'. The dash-dot line shown in figure (a) represents the boundary for statically stable, below which the gas is heavy at bottom.}
	\label{fig:comp_finger_145}
\end{figure}

Figure \ref{fig:comp_finger_145} plots $\rm Ra_T$, ${\rm d}{\rm Ra_T}/{\rm d}{\rm Ra_S}$ and $k$ for FC, FV and BV' at neutral stability in the fingering regime. First examine the influences of variable properties through comparing FC and FV. It is evident that the variable properties have significant effects on the stability. As $|{\rm Ra_S}|$ is increased and larger than $2\times 10^{5}$, the variable properties make the neutral curve bend downward. The stability is actually weakened since the stable region is reduced in size (Figs. \ref{fig:comp_finger_145}(a)). Besides, it can be seen in Fig. \ref{fig:comp_finger_145}(c), the wave number grows more quickly than that under constant properties. 

\begin{figure}
	\centering
	\includegraphics[width=13cm]{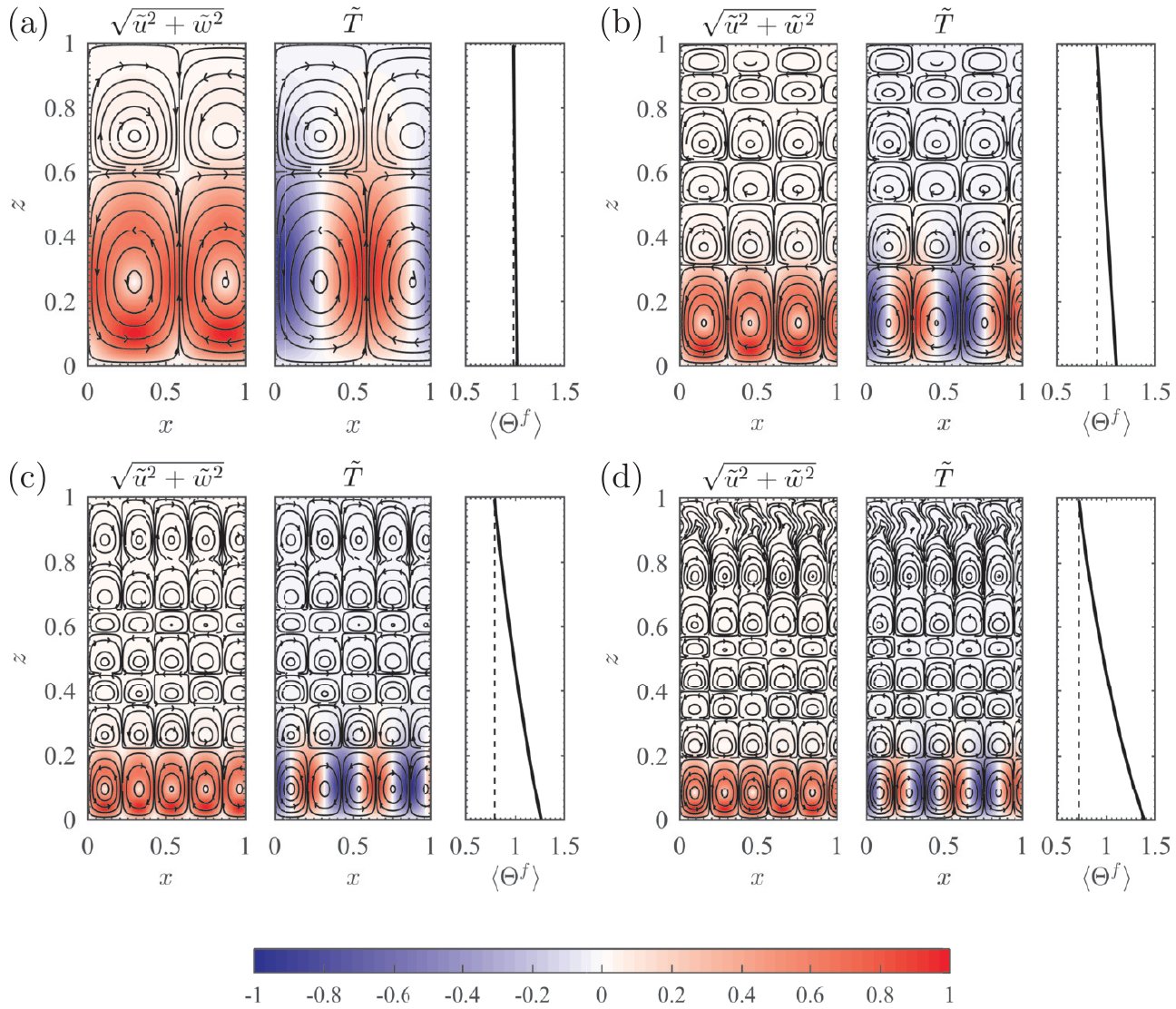}
	\caption{The description is the same with Fig. \ref{fig:fields_fin_1}, where the role of FC is replaced by FV, and $\tau^{\ast-1}$ by $\Theta^f$. $\Theta^f$ is defined by Eq. (\ref{eq:theta_fin}).}
	\label{fig:fields_fin_4}
\end{figure}
Figure \ref{fig:fields_fin_4} shows the profiles of velocity and temperature perturbations, along with flow structure, obtained from FV for different $\rm Ra_S$ chosen according to Fig. \ref{fig:fields_fin_1}. It is shown that the physical space is layered by counter-rotating vortices, and the flow is dominated by a specific layer, that is attributed to penetrative instability. However, the flow patterns in Fig. \ref{fig:fields_fin_4} are different from those in Fig. \ref{fig:fields_fin_1}. First, the unstable layer changes from the top one to the bottom one, since the intensity of flow is dominated by the bottom layer. Second, some irregular variations are detected along the penetrative depth. In the penetrative instability illustrated in Fig. \ref{fig:fields_fin_1}, the size and intensity of vortices decrease gradually along the penetrative direction (downward). But when variable properties are present, the variations in vortices along the penetrative direction (upward) become irregular. Some unexpected complex flow patterns occur near the top wall when the $|{\rm Ra_S}|$ is large enough (see Fig. \ref{fig:fields_fin_4}(c) and \ref{fig:fields_fin_4}(d)).

Now the question is how to link the variations in physical properties with the penetrative instability. As an example, the physical properties and state variables as functions of $z$ for ${\rm Ra_S} = -1\times10^7$ and ${\rm Ra_T}=-1.03\times10^8$ (on the stability boundary) has been shown in Fig. \ref{fig:variable_properties_fin}. Since all physical properties and state variables vary nonlinearly with $z$, it is necessary to develop a comprehensive parameter based on the underlying mechanism, instead of using a single physical property, to interpret the irregular penetrative instability.

The energy sustaining the growth of small perturbations comes from the potential energy stored in the component that is unstably distributed \cite{huppert_1981}. But the stably distributed component hinders the release of potential energy. In the fingering regime, concentration is the unstable component while temperature is the stable component. This suggests using the ratio of potential energy as an indicator for the penetrative instability, leading to
\begin{equation}\label{eq:Rrho_fin}
R_\rho^f(z) = \frac{\kappa_r\nabla_c}{\beta_r(\nabla_T-\nabla_{ad,r})}\frac{\kappa c_z}{\beta T_z}=\frac{{\rm Ra_S}}{{\rm Ra_T}}\frac{\kappa \bar c_z}{\beta \bar T_z},
\end{equation}
which is actually the reciprocal of the density ratio widely used in the unbounded DDC \cite{radko2013double}. Eq. (\ref{eq:Rrho_fin}) can also be understood with the help of Fig. \ref{fig:ill_fin_rho}. Assuming $D$, $D_T$ and $\kappa \bar c_z$ are constant, if $\beta \bar T_z$ is increased, after perturbing a fluid parcel downward by $\zeta$, it will take a longer time to eliminate the stabling effect of temperature. Meanwhile, the ongoing concentration diffusion decreases its density difference with surroundings. As a result, the amplification of small perturbation is weakened. It is equivalent to reducing $\kappa \bar c_z$ while maintaining $\beta \bar T_z$. Above descriptions give a clear picture for the mechanism behind Eq. (\ref{eq:Rrho_fin}).

%

Moreover, the last section showed that it has been concluded that the intensity of flow is also inversely proportional to the actual diffusivity ratio, defined as
\begin{equation}\label{eq:tau_fin}
{\tau ^ * } (z)= \tau \frac{D}{{{D_T}}},
\end{equation}
where $D_T=\lambda/(\bar\rho c_p)$ is the non-dimensional thermal diffusivity. Combining Eqs. (\ref{eq:Rrho_fin}) and (\ref{eq:tau_fin}) yields a comprehensive parameter $\Theta^f$ given by
\begin{equation}\label{eq:theta_fin}
\Theta^f(z) =\frac{ R_{\rho}^f}{\tau^*}=\frac{{\rm Ra_S}}{{\tau {{\rm Ra_T}}}}\frac{{{D_T}\kappa {\bar c_z}}}{{D\beta {\bar T_z}}}.
\end{equation}
The vertical distributions of $\Theta^f$ are plotted in Fig. \ref{fig:fields_fin_4}. Results show $\Theta^f$ is a good indicator for the penetrative instability. The unstable bottom layer is always a region with large $\Theta^f$. The non-uniformity of $\Theta^f$ measures the degree of layering, which is evidenced by the fact that in Fig. \ref{fig:fields_fin_4} from (a) to (d), the non-uniformity of $\Theta^f$ and the number of layers increase gradually and simultaneously. In addition, when all physical properties are constant, $R_{\rho}^f={\rm Ra_T}/{\rm Ra_S}$ and $\Theta^f=R_{\rho}^f/\tau^\ast\propto\tau^{\ast-1}$. $\Theta^f$ degenerates to the indicator proposed in the last section multiplied by a $z$-independent coefficient.

Besides, it is confirmed by Fig. \ref{fig:comp_finger_145} that the results from BV' are in good agreement with FV, implying the modified BA captures the underlying mechanism of instability and can be used to simplify the original equations.

It is shown in this section that in the fingering regime, when $|{\rm Ra_S}|$ is large enough, the base-state density stratification and variations in physical properties become important, and the former makes the BA break down. The fluid close to the lower boundary is more unstable leading to a penetrative instability. The modified BA incorporating variable properties can be used to simplify the original equations with good accuracy. Besides, a comprehensive parameter $\Theta^f$ is developed to measure the non-uniformity in physical properties. 


In fact, there are some previous studies concerning DDC with variable properties or nonlinear state variables. For example, \citet{walton_1982} and \citet{Balmforth1998} study the onset of DDC with nonlinear concentration distributions. \citet{TANNY19951683} investigate influences of temperature-dependent viscosity and diffusivity on the onset of DDC. To our knowledge, the onset of DDC with variations in all physical properties, plus nonlinear state variables, is investigated here for the first time. Eq. (\ref{eq:theta_fin}) provides a simple but effective way to measure these non-uniformities and nonlinearities. In order to give more insights into $\Theta$, the relationship between $\Theta$ and qualitative deviations of BC from FV is studied in section \ref{sec:fd}.

\section{Oscillatory regime}\label{osi_Bou}
\subsection{The applicability of the Boussinesq approximation}\label{sec:BA}
\begin{figure}	
	\centering
	\includegraphics[width=13cm]{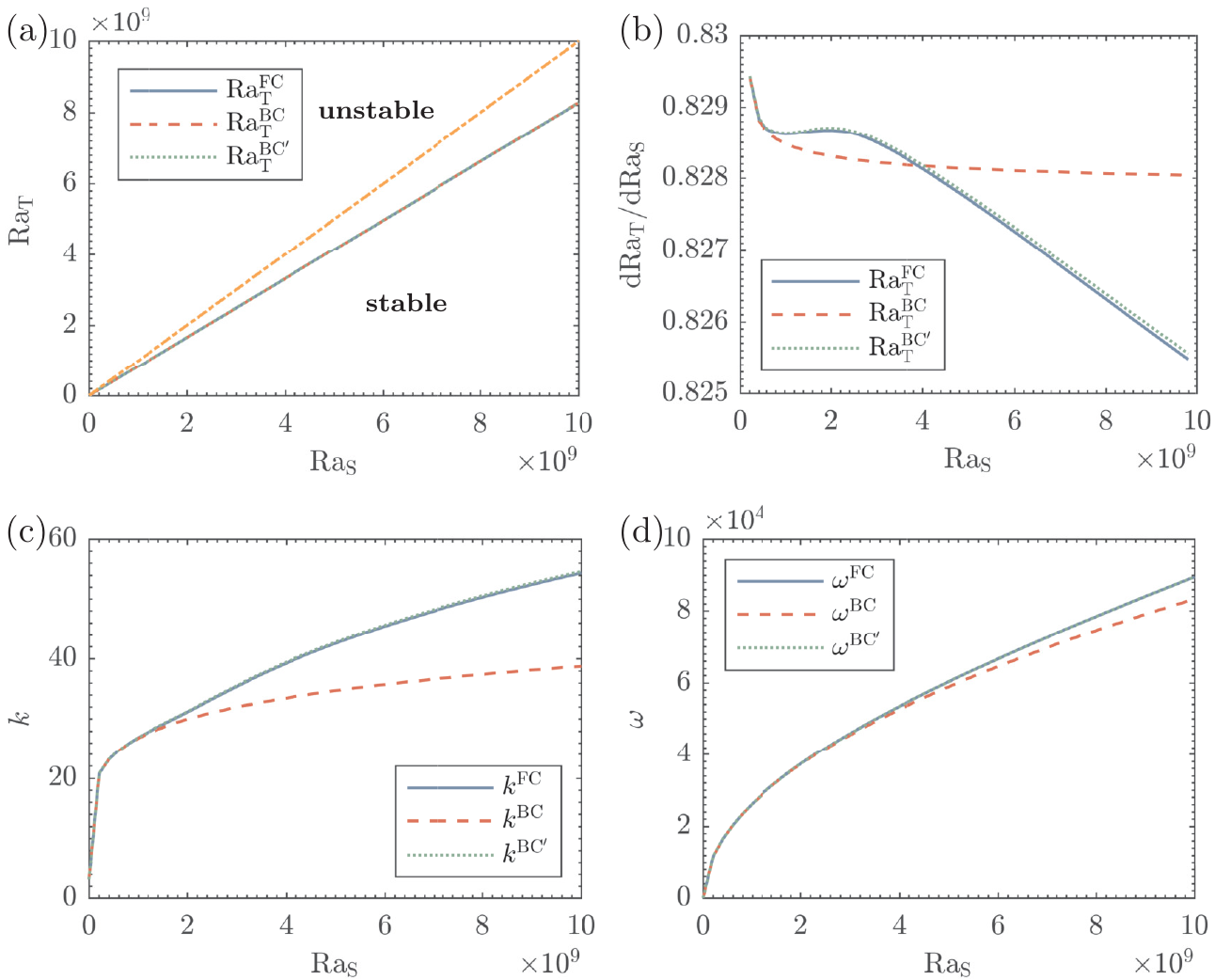}
	\caption{Comparisons among FC, BC and BC' in the oscillatory regime. (a): Neutral curves satisfying ${\rm Re}(\sigma)=0$, (b): slopes of the neutral curves, (c): critical wave numbers, (d): angular frequencies, with solid lines for FC, dashed lines for BC and dotted lines for BC'. The dash-dot line shown in figure (a) represents the boundary for statically stable, below which the gas is heavy at bottom.}
	\label{fig:comp_osi_123}
\end{figure}

As in the previous chapter, FC, BC and BC' are discussed here. At neutral stability in the oscillatory regime (${\rm Ra_T}>0$, ${\rm Ra_S}>0$), the growth rate is a pure imaginary. The imaginary part of the growth rate $\omega = {\rm Im}(\sigma)$, is the angular frequency, which also depends on $\rm Ra_S$. In Fig. \ref{fig:comp_osi_123}, in order to study the applicability of the BA, the thermal Rayleigh number, the horizontal wave number $k$, and the angular frequency $\omega$ are plotted as functions of $\rm Ra_S$ for FC, BC and BC'. The range of $\rm Ra_S$ shown here is $[0,1\times10^{10}]$.

As shown in Fig. \ref{fig:comp_osi_123}(a), the oscillatory instability is triggered under a bottom-heavy configuration. The relative differences among the three curves are too small to separate. Figure \ref{fig:comp_osi_123}(b) shows $\mathrm{d}{\rm Ra_T}/\mathrm{d}{\rm Ra_S}$ and as $\rm Ra_S$ is increased, the deviations resulting from the BA appear here for ${\rm Ra_S} \approx6\times10^8$. The BA cannot reproduce the non-monotonic variations of $\rm Ra_T$ and, $k$ and $\omega$ are underestimated. The differences in $k$ and $\omega$ are not significant until ${\rm Ra_S} \approx1.5\times10^9$ and ${\rm Ra_S} \approx3\times10^9$, respectively
(see Fig. \ref{fig:comp_osi_123}(c) and \ref{fig:comp_osi_123}(d)). 

\begin{figure}
	\centering
	\includegraphics[width=13cm]{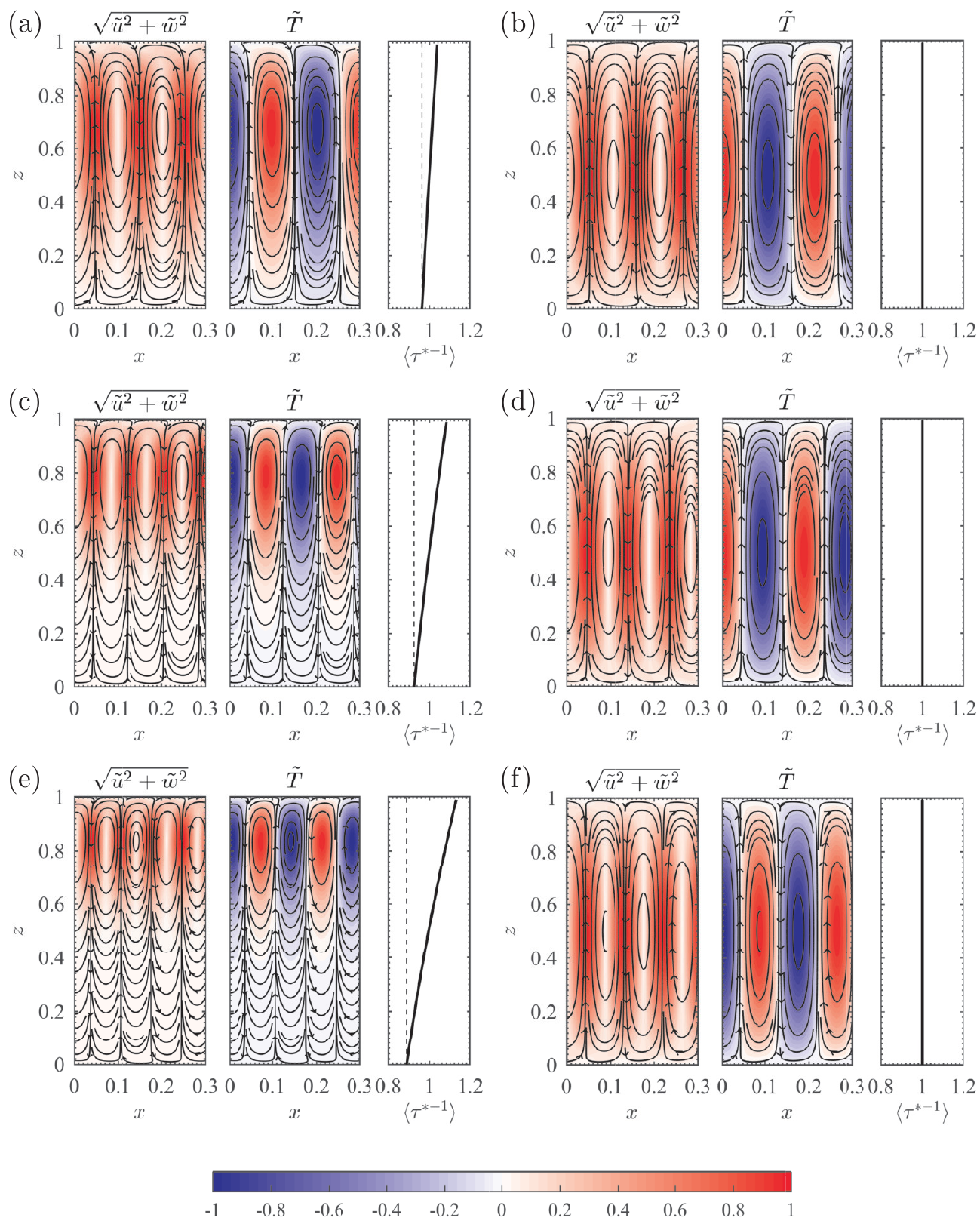}
	\caption{See the caption of Fig. \ref{fig:fields_fin_2} for the description of contours and streamlines, with (a) \& (b): ${\rm Ra_S} = 1.84\times10^9$, (c) \& (d): ${\rm Ra_S} = 3.88\times10^9$ and (e) \& (f): ${\rm Ra_S} = 5.92\times10^9$. (a), (c) and (e) are obtained from FC. (b), (d) and (f) are obtained from BC. The range of $x$ presented is $[0,0.3]$. $\tau^{*}$ is the actual thermal diffusivity given by Eq. (\ref{eq:tau_sat}). $\langle\tau^{*-1}\rangle = \tau^{*-1}/\tau^{*-1}_{m}$, with $\tau^{*-1}_{m}$ denoting its mean value. To make the vertical non-uniformity of $\langle\tau^{*-1}\rangle$ clear, additional dashed lines representing the minimum value of  $\langle\tau^{*-1}\rangle$ are plotted.}
	\label{fig:comp_fields_12}
\end{figure}
Figure \ref{fig:comp_fields_12} shows the contours of velocity and temperature perturbations, along with flow structure obtained from FC and BC for ${\rm Ra_S} = 1.84\times10^9$, ${\rm Ra_S} = 3.88\times10^9$ and ${\rm Ra_S} = 5.92\times10^9$. In contrast to the fingering regime, interesting off-center structures are observed. For all figures in BC, the vortices are always centered on $z=0.5$. However, the centers of vortices in FC move upward as $\rm Ra_S$ is increased. As a result, the flow in the upper part is stronger than that in the lower part. The upward shift of vortices' centers increases with $\rm Ra_S$. For Figs. \ref{fig:comp_fields_12}(a), \ref{fig:comp_fields_12}(c) and \ref{fig:comp_fields_12}(e), the centers are approximately located on $z = 0.68$, 0.78 and 0.85, respectively. Besides, an increase in $k$ for FC is indicated by Fig. \ref{fig:comp_osi_123}(c).

\begin{figure}
	\centering
	\includegraphics[width=7.5cm]{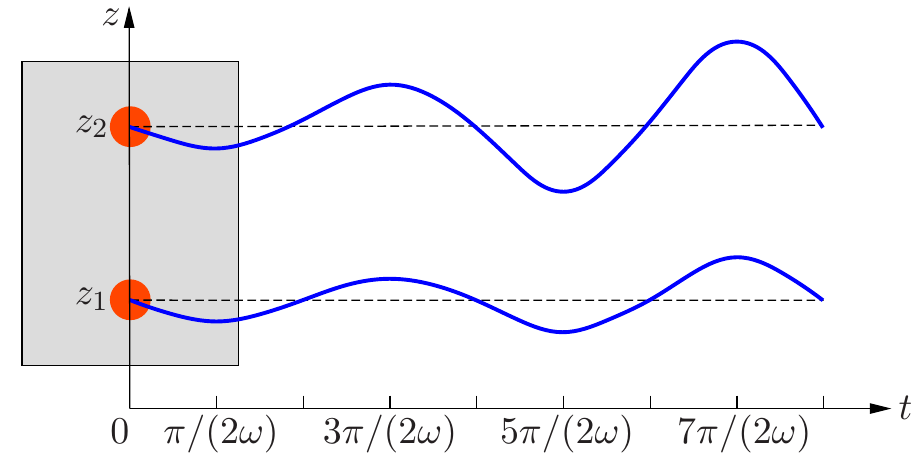}
	\caption{Illustration of the effect of base-state density stratification in the oscillatory regime.}
	\label{fig:ill_osi_rho}
\end{figure}

According to the discussions in section \ref{sec:fin_bou}, the off-center structure should be caused by the base-state density stratification. This interpretation is confirmed by Fig. \ref{fig:comp_osi_123}, where the results from BC' are in good agreement with those from FC. Now the question is why does the upward-decreasing base-state density causes the off-center structure? To answer this question, consider the actual dimensional thermal diffusivity defined by Eq. (\ref{eq:real_DT}), which increases upward due to the reciprocal of $\bar\rho$. Figure \ref{fig:ill_osi_rho}, based on the classical illustration of oscillatory DDC \cite*{radko2013double,garaud2017}, represents a thought experiment helping one to understand the mechanism. Consider two parcels of fluid, located at $z_1$ and $z_2$, with $z_1<z_2$. Perturb them by displacing downward the same distance at $t = \pi/(2\omega)$. At new positions, they are cooler with lower concentration than their surroundings. Since $\tau<1$, they adjust their temperature quickly but not their concentration. Then, they become lighter than their surroundings, and move upward driven by buoyancy. However, the upper fluid parcel, with a large $D_T^\ast$, adjusts its temperature more efficiently than the lower one, hence driven by a stronger buoyancy. As a result, the upper fluid parcel overshoots its original equilibrium position more greatly than the lower one (see Fig. \ref{fig:ill_osi_rho} at $t = 3\pi/(2\omega)$). Above their original positions, they again become heavier than their surroundings due to heat loss, and the gravity drives them back. Based on aforementioned mechanism, the downward driving force for the upper fluid parcel is larger than the lower one. The upper fluid parcel again overshoots more greatly than the lower one (see Fig. \ref{fig:ill_osi_rho} at $t = 5\pi/(2\omega)$). Above process repeats and eventually, the perturbation of the upper parcel amplifies more greatly than the lower one.

Ratio given by Eq. (\ref{eq:tau_sat}) is still considered as the actual diffusivity ratio. Plot $\tau^{\ast-1}$ against $z$ in Fig. \ref{fig:comp_fields_12}.  It is demonstrated that the amplitude of perturbation is proportional to $\tau^{\ast-1}$. And the upward shift of vortex center increases with the gradient of $\tau^{\ast-1}$. 

\subsection{The effects of variable properties}
\begin{figure}
	\centering
	\includegraphics[width=13cm]{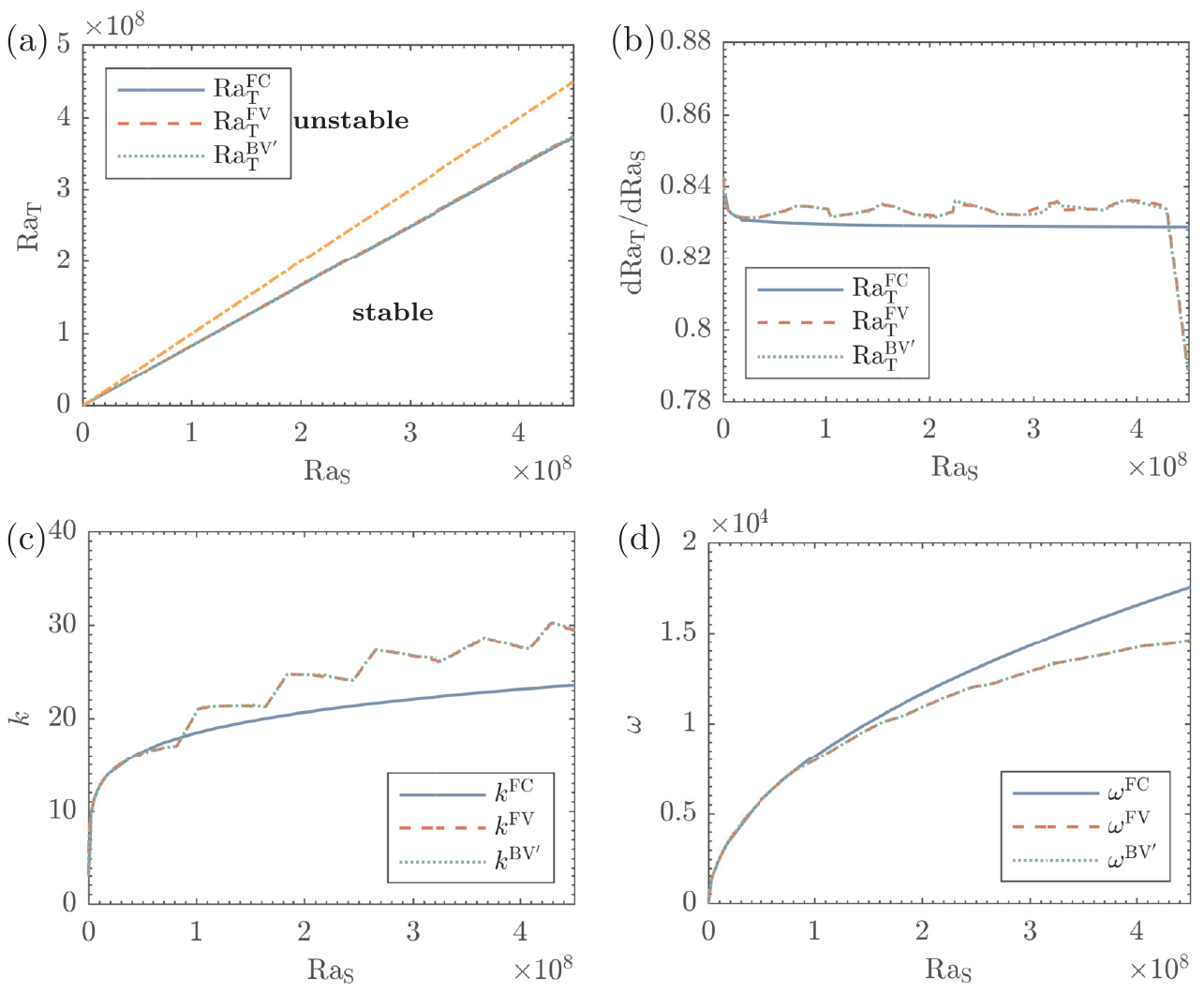}
	\caption{Comparisons among FC, FV and BV' to present the influence of variable properties on neutral stability in the oscillatory regime. (a): Neutral curves satisfying ${\rm Re}(\sigma)=0$, (b): slopes of the neutral curves, (c): critical wave numbers, (d): angular frequencies, with solid lines for FC, dashed lines for FV and dotted lines for BV'. The dash-dot line shown in figure (a) represents the boundary for statically stable, below which the gas is heavy at bottom.}
	\label{fig:comp_osi_145}
\end{figure}

Figure \ref{fig:comp_osi_145} plots $\rm Ra_T$, $k$ and $\omega$ for FC, FV and BV' in the oscillatory regime. The range of $\rm Ra_S$ shown here is $[0, 4.5\times10^8]$. First discuss the influences of variable properties through comparing FC and FV. After the variations in physical properties are considered, the convection is still triggered under a bottom-heavy configuration (see Fig. \ref{fig:comp_osi_145}(a)). However, as $\rm Ra_S$ is increased, the variations in physical properties become more significant, the differences between FC and FV increase gradually and become obvious when ${\rm Ra_S}\ge2.0\times10^7$ (see Figs. \ref{fig:comp_osi_145}(b), \ref{fig:comp_osi_145}(c) and \ref{fig:comp_osi_145}(d)). Besides, the influence of variable properties is not monotonic, and noticeable variations are observed. 

As for the accuracy of BV', Fig. \ref{fig:comp_osi_145} demonstrates that the dashed curves in FV and the dotted curves in BV' coincide, except for the unavoidable numerical differences.  We conclude that in the oscillatory regime when variable properties are introduced, BV' is accurate and can be used as an approximation of FV. 

\begin{figure}
	\centering
	\includegraphics[width=13cm]{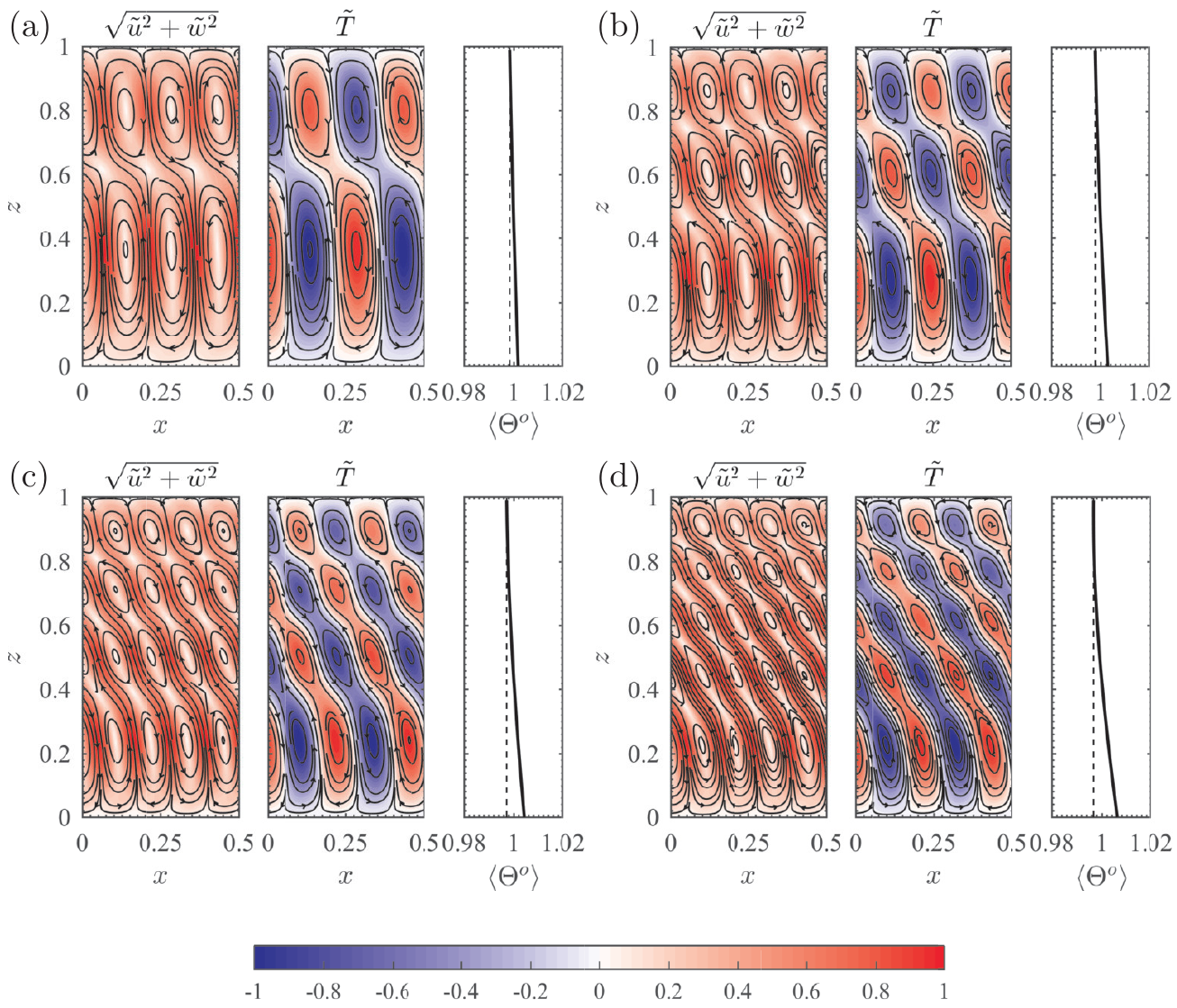}
	\caption{See the caption of Fig. \ref{fig:fields_fin_2} for the description of contours and streamlines, with (a):  ${\rm Ra_S} = 1.22\times10^8$, (b): ${\rm Ra_S} = 1.84\times10^8$, (c): ${\rm Ra_S} = 2.65\times10^8$ and (d): ${\rm Ra_S} = 3.67\times10^8$. The range of $x$ presented is $[0,0.5]$. $\Theta^o$ is given by Eq. (\ref{eq:theta_osi}), and $\langle\Theta^o\rangle = \Theta^o/\Theta^o_{m}$, with $\Theta_{m}^o$ denoting its mean value. To make the vertical distribution of $\langle\Theta^o\rangle$ clear, additional dashed lines representing the minimum value of  $\langle\Theta^o\rangle$ are plotted.}
	\label{fig:fields_osi_4}
\end{figure}
\begin{figure}
	\centering
	\includegraphics[width=8cm]{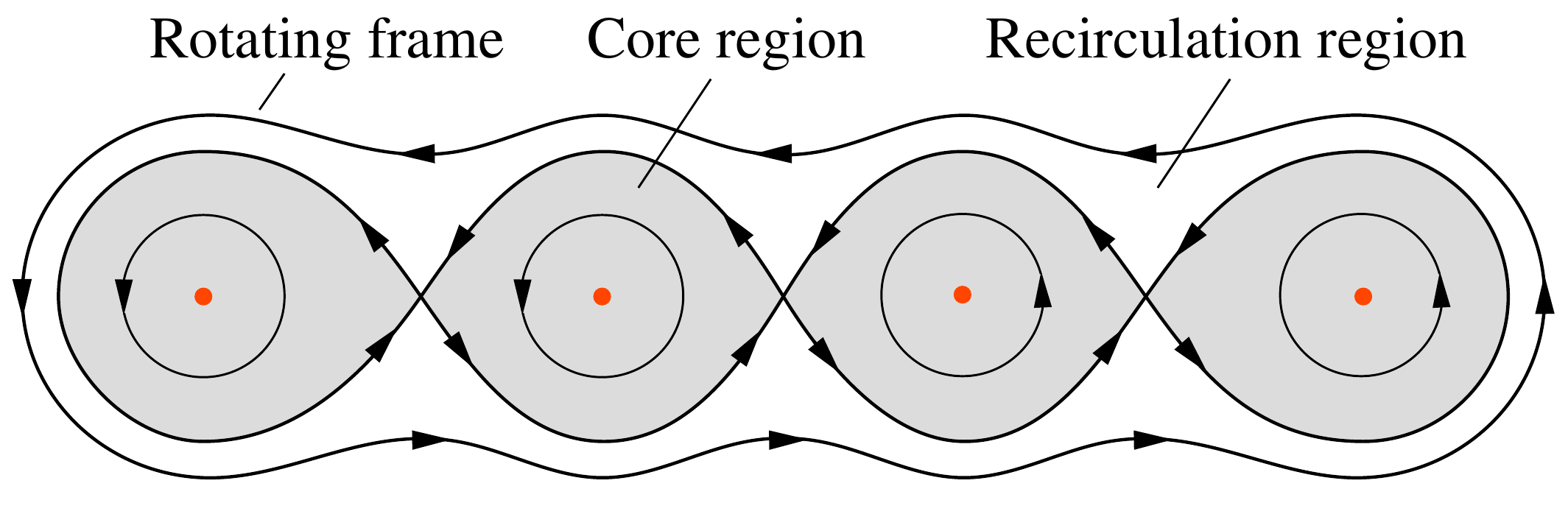}
	\caption{Illustration of the cat's eye pattern in Fig. \ref{fig:fields_osi_4}.}
	\label{fig:vortex_struc_eye}
\end{figure}

Figure \ref{fig:fields_osi_4} presents the perturbation fields and flow structure for different $\rm Ra_S$ when the role of variable properties becomes important. Instead of penetrative instability, the vertical vortex tilts and internally breaks up into several corotating vortices. Such an intriguing flow field is called the cat's eye pattern, which is illustrated in Fig. \ref{fig:vortex_struc_eye}. The cat's eye pattern comprises a rotating frame, a core region and a recirculation region between the two. The core region comprises an array of corotating vortices, around which the fluid in recirculation region travels \cite{Leweke2016}. In addition, the number of corotating vortices is proportional to $\rm Ra_S$. And the size and intensity of the vortices decrease upward. Since the onset of DDC in a near-critical gas is studied for the first time, the above phenomenon has not been reported before.

As an example, the physical properties and state variables as functions of $z$ for ${\rm Ra_S} = 3.67\times10^8$ and ${\rm Ra_T}=3.06\times10^8$ (on the stability boundary) has been shown and discussed eailier (see Fig. \ref{fig:variable_properties_osi}). In section \ref{sec:vp_fin}, combining Eqs. (\ref{eq:Rrho_fin}) and (\ref{eq:tau_fin}), a comprehensive parameter $R_{\rho}^f$ given by Eq. (\ref{eq:theta_fin}) is proposed to interpret the influence of the variations in physical properties based on an energy storage and release mechanism. However, in the oscillatory regime, the energy sustaining the amplification of small perturbation is provided by temperature and hindered by concentration. Therefore, $R_\rho$ in the oscillatory regime, denoted by $R_\rho^o$, is given by
\begin{equation}\label{eq:Rrho_osi}
R_\rho^o(z) = \frac{1}{R_\rho^f(z)}=\frac{{\rm Ra_T}}{{\rm Ra_S}}\frac{\beta \bar T_z}{\kappa \bar c_z}.
\end{equation}
The mechanism behind this equation can also be interpreted through the thought experiment shown in Fig. \ref{fig:ill_osi_rho}. We assume $D$ and $D_T$ are constant and consider the process of a fluid parcel downward leaving its original equilibrium position and return. During the whole process, its temperature and concentration will increase. Then at its initial position, the buoyancy force governing its amplification, is provided by temperature but compensated by concentration. In Eq. (\ref{eq:Rrho_osi}), if $\beta T_z$ is increased while $\kappa c_z$ remains unchanged, then the buoyancy will increase. On the contrary, if $\kappa c_z$ is increased while $\beta T_z$ remains unchanged, then the buoyancy will decrease. In summary, Eq. (\ref{eq:Rrho_osi}) reflects the competition between the two factors governing the amplification of small perturbation. 

Based on the arguments in section \ref{sec:vp_fin}, combine Eqs. (\ref{eq:Rrho_osi}) and (\ref{eq:tau_fin}) to obtain a comprehensive parameter $\Theta^o$:
\begin{equation}\label{eq:theta_osi}
\Theta^o(z) =\frac{ R_{\rho}^o}{\tau^*}=\frac{{\rm Ra_T}}{{\tau {{\rm Ra_S}}}}\frac{{{D_T}\beta {\bar T_z}}}{{D\kappa {\bar c_z}}}.
\end{equation}
The vertical distributions of $\Theta^o$ are plotted in Fig. \ref{fig:fields_osi_4}. Results show $\Theta^o$ is a good indicator for the cat's eye pattern. First, $\Theta^o$ is proportional to the size and intensity of the vortices. Second, the non-uniformity of $\Theta^o$ measures the number of inner vortices, which is evidenced by the fact that in Fig. \ref{fig:fields_osi_4} from (a) to (d), the non-uniformity of $\Theta^o$ and the number of vortices in every singe frame increase simultaneously. Thirdly, when all physical properties are constant, $R_{\rho}^o={\rm Ra_T}/{\rm Ra_S}$ and $\Theta^o=R_{\rho}^o/\tau^\ast\propto\tau^{\ast-1}$. $\Theta^o$ actually degenerates to the indicator proposed in section \ref{sec:BA} multiplied by a $z$-independent coefficient.

This section suggests the base-state density stratification and variable properties also have significant effects on the onset of convection in the oscillatory regime. The omitting of base-state density stratification leads to the break-down of the BA. The modified BA still works well here. The onset of convection features a cat's eye pattern caused by the nonlinearly distributed physical properties, instead of the penetrative instability detected in the fingering regime. The modified BA as proposed in the last section still holds in the oscillatory regime. The comprehensive parameter, $\Theta^o$, is also derived in the oscillatory regime based on the previous arguments. 

\section{Further discussions of $\Theta$}\label{sec:fd}
So far, a comprehensive parameter $\Theta$ has been defined, whose non-uniformity is proportional to effects of variable properties. In this section, a quantitative analysis of the influence of its non-uniformity on the DDC is performed based on BC and FV. Afterward, the relative contributions of the two controlling factors in the non-uniformity of $\Theta$ are discussed. Because only two cases are involved, we use $\Theta$ to denote $\Theta$ in FV, and $\Theta_{BC}$ to $\Theta$ in BC.

A question naturally arises is how to quantify the non-uniformity of $\Theta$. This requires one to define a reference, according to which the non-uniformity is measured. In this study, it is natural to use $\Theta _{\rm BC}$ as a reference state. Mathematically, the non-uniformity of $\Theta$, denoted by $A_{\Theta}$, can be expressed by
\begin{equation}\label{eq:A_theta}
A_{\Theta} = d\int_0^1 |{\Theta  - {\Theta _{\rm BC}}}| {\rm d} z.
\end{equation}
The calculations for $A_{\Theta}$ were performed for both the fingering regime and the oscillatory regime, and the results are denoted by $A_{\Theta}^f$ and $A_{\Theta}^o$, respectively. 

\begin{figure}
	\centering
	\includegraphics[width=13cm]{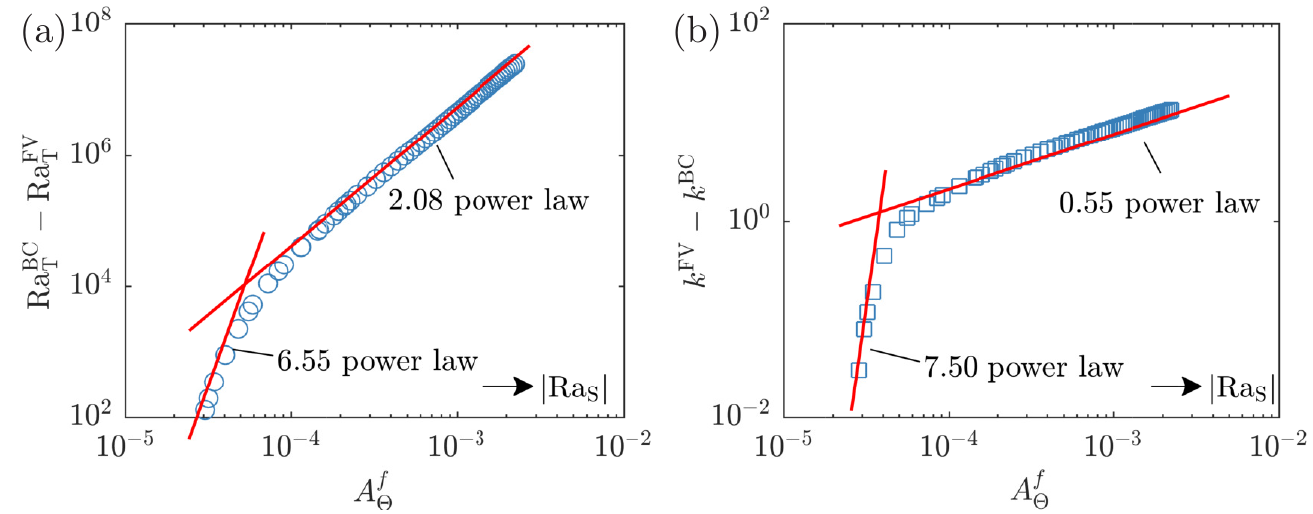}
	\caption{(Log-log plot) The variations of absolute differences in (a) thermal Rayleigh numbers and  (b) wave numbers at neutral states obtained from BC and FV, versus $A_{\theta}^f$. The four power laws are represented by the solid lines. The absolute value of $\rm Ra_S$, denoted by $|{\rm Ra_S}|$, increases along the direction represented by the arrow.}
	\label{fig:fin_theta}
\end{figure}

First focus on the fingering regime. Presented in Fig. \ref{fig:fin_theta} is the variations of absolute differences in thermal Rayleigh numbers and wave numbers at neutral states obtained from BC and FV. The differences are defined by $({\rm Ra_T}^{\rm BC}-{\rm Ra_T}^{\rm FV})$ and $(k^{\rm FV}-k^{\rm BC})$ respectively to produce positive values. As mentioned earlier, the deviations of FV from BC are proportional to the variations in physical properties. This conclusion is clearly shown in Fig. \ref{fig:fin_theta}, where both $({\rm Ra_T}^{\rm BC}-{\rm Ra_T}^{\rm FV})$ and $(k^{\rm FV}-k^{\rm BC})$ increase along with $A_{\theta}^f$. Remarkably, four power laws are identified by fitting (represented by solid lines). For $({\rm Ra_T}^{\rm BC}-{\rm Ra_T}^{\rm FV})$, in the high $A_{\theta}^f$ region, the data follows a 2.08 power law, whereas in the low $A_{\theta}^f$ region, the data follows a 6.55 power law. As to $(k^{\rm FV}-k^{\rm BC})$, a 0.55 power law and a 7.50 one are identified in the high $A_{\theta}^f$ region and the low $A_{\theta}^f$ region, respectively. Mathematically, the power laws are given by
\begin{eqnarray}
({\rm Ra_T}^{\rm BC}-{\rm Ra_T}^{\rm FV})&\sim&\left\{
\begin{array}{ll}
(A_{\theta}^f)^{6.55}, & A_{\theta}^f< 4.0\times10^{-5}, \\[2pt]
(A_{\theta}^f)^{2.08}, & A_{\theta}^f>1.5\times10^{-4},
\end{array} \right.\label{eq:pl_1}\\
(k^{\rm FV}-k^{\rm BC})&\sim&\left\{
\begin{array}{ll}
(A_{\theta}^f)^{7.50}, & A_{\theta}^f< 4.0\times10^{-5}, \\[2pt]
(A_{\theta}^f)^{0.55}, & A_{\theta}^f>1.5\times10^{-4}.
\end{array} \right.
\end{eqnarray}
The range $4.0\times10^{-5}<A_{\theta}^f<1.5\times10^{-4}$ defines a transition between the two power laws. 

\begin{figure}
	\centering
	\includegraphics[width=13cm]{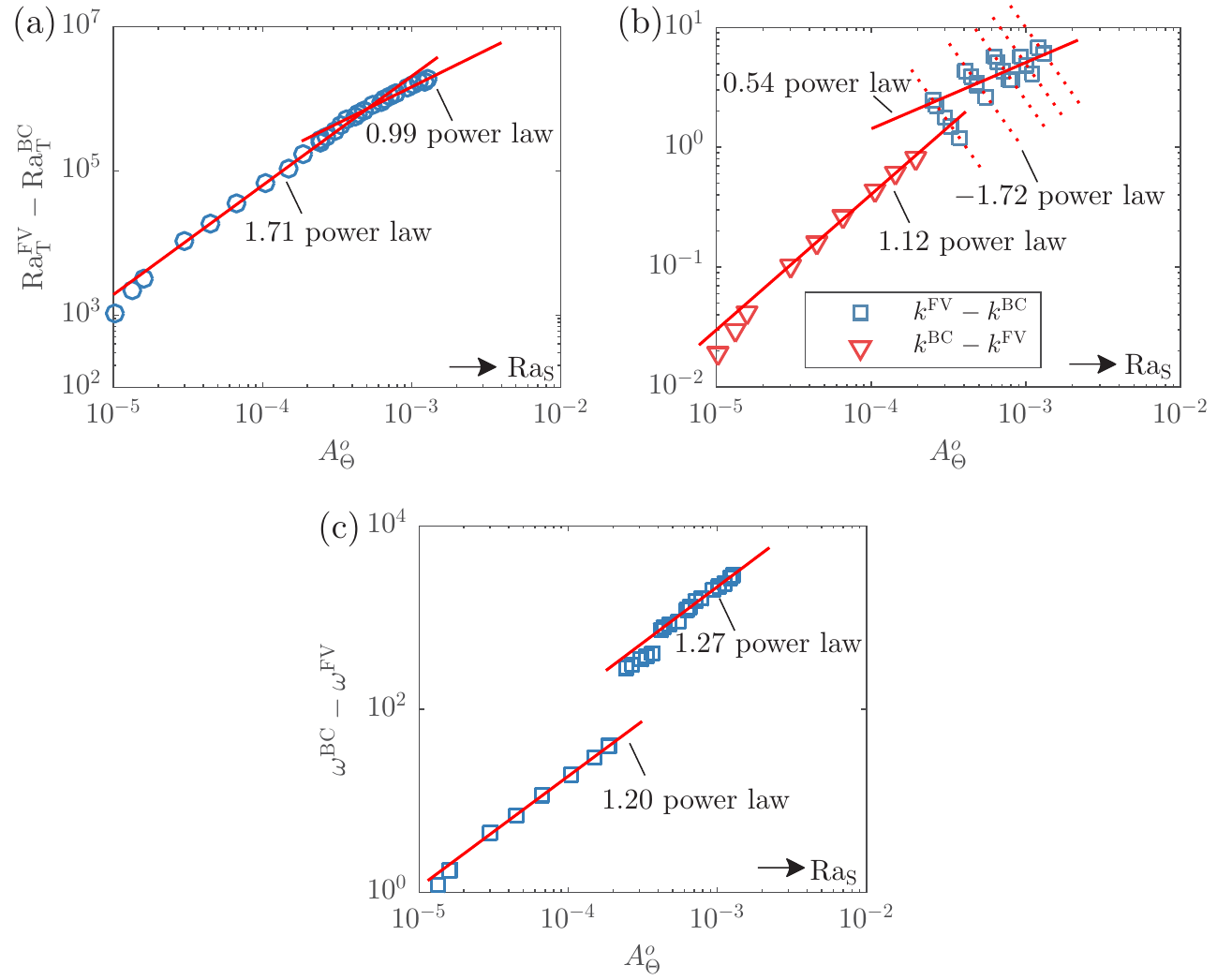}
	\caption{(Log-log plot) The variations of absolute differences in (a) thermal Rayleigh numbers,  (b) wave numbers and (c) angular frequencies at neutral states obtained from BC and FV, versus $A_{\theta}^o$. Several power laws are identified and represented by solid and dotted lines. $\rm Ra_S$ increases along the direction represented by the arrow.}
	\label{fig:osi_theta}
\end{figure}
For the oscillatory regime, Fig. \ref{fig:osi_theta} presents absolute differences in thermal Rayleigh numbers (${\rm Ra_T}^{\rm FV}-{\rm Ra_T}^{\rm BC}$), wave numbers ($k^{\rm FV}-k^{\rm BC}$) or ($k^{\rm BC}-k^{\rm FV}$), and angular frequencies ($\omega^{\rm BC}-\omega^{\rm FV}$) against $A_{\theta}^o$. These absolute differences are defined to produce positives values. Obvious fluctuations are shown in figures	\ref{fig:osi_theta}(b) and \ref{fig:osi_theta}(c) when $A_{\theta}^o>2\times 10^{-4}$, as introduced in Fig. \ref{fig:comp_osi_145}. These fluctuations bring difficulties in studying quantitative relationships between $A_{\theta}^o$ and absolute differences. Therefore, mean values in each fluctuation are calculated and used in fitting. Several power laws are identified and represented by solid lines in Fig. \ref{fig:osi_theta}. In the monotonic region, namely $A_{\theta}^o<2.0\times10^{-4}$, these power laws hold:
\begin{eqnarray}
({\rm Ra_T}^{\rm FV}-{\rm Ra_T}^{\rm BC})&\sim& (A_{\theta}^o)^{1.71},\nonumber\\
(k^{\rm BC}-k^{\rm FV})&\sim& (A_{\theta}^o)^{1.12},\\
(\omega^{\rm BC}-\omega^{\rm FV})&\sim& (A_{\theta}^o)^{1.20}. \nonumber
\end{eqnarray}
In the fluctuation region, namely $A_{\theta}^o>2.0\times10^{-4}$, the power laws for mean value are given by
\begin{eqnarray}
({\rm Ra_T}^{\rm FV}-{\rm Ra_T}^{\rm BC})_m&\sim& (A_{\theta}^o)^{0.99},  A_{\theta}^o>4.0\times10^{-4},\nonumber\\
(k^{\rm FV}-k^{\rm BC})_m&\sim& (A_{\theta}^o)^{0.54},  A_{\theta}^o>4.0\times10^{-4},\\
(\omega^{\rm BC}-\omega^{\rm FV})_m&\sim& (A_{\theta}^o)^{1.27},  A_{\theta}^o>4.0\times10^{-4},\nonumber
\end{eqnarray}
with the subscript $m$ denoting a mean value. The range $2.0\times10^{-4}<A_{\theta}^o<4.0\times10^{-4}$, in which the first fluctuation occurs, defines a transition between the two power laws. In addition, it is interesting to find that the ($k^{\rm FV}-k^{\rm BC}$) in each fluctuation obey a $-1.75$ power law, namely
\begin{equation}\label{eq:pl_end}
(k^{\rm FV}-k^{\rm BC})\sim(A_{\theta}^o)^{-1.75},
\end{equation}
which has been shown in Fig. \ref{fig:osi_theta}(b) by dotted lines. 

\section{Conclusions}\label{sec:con}
The physical model considered in this study is a layer of near-critical binary gas with infinite extends bounded by two horizontal rigid walls, with Dirichlet boundary conditions for both temperature and concentration. The onset of DDC in it due to vertical temperature and concentration gradients is numerically investigated. The non-dimensional governing system contains seven parameters, in which the definition of thermal Rayleigh number is modified based on previous studies on near-critical RB problem. A numerical LSA is performed, and equations for small perturbations are solved based on a finite-difference discretization. Two problems arise: the applicability of the BA, and the influences of the variations in physical properties on DDC. To this end, five cases of mathematical models are shown based on different degrees of simplification. 

The applicability of the BA is first studied. It is revealed that due to strong base-state density stratification, the thermal diffusivity in the upper region is magnified. As a result, regular penetrative instability (Fig. \ref{fig:fields_fin_1}) and off-center structures (Fig. \ref{fig:comp_fields_12})  are identified in the fingering regime and the oscillatory regime, respectively. However, since the base-state density stratification is omitted in the BA, the flows are always featured by a vertical single vortex centered on the middle (Figs. \ref{fig:fields_fin_2} and \ref{fig:comp_fields_12}). Besides, a modified BA is proposed adding the effect of base-state density stratification. In the fingering regime, the modification brings big improvements. And the accuracy of the modified BA is excellent in the oscillatory regime. 

The variations in physical properties have significant effects on DDC. Irregular penetrative instability (Fig. \ref{fig:fields_fin_4}) and cat's eye pattern (Figs. \ref{fig:fields_osi_4} and \ref{fig:vortex_struc_eye}) are thus identified in the fingering and oscillatory regime, respectively. For the $\rm CO_2-C_2H_6$ mixture at the reference state, these effects become obvious when ${\rm Ra_S}<-2\times10^{5}$ in the fingering regime and ${\rm Ra_S}>2\times10^{7}$ in the oscillatory regime. A comprehensive parameter $\Theta$ based on an energy storage and release mechanism is proposed in Eqs. (\ref{eq:theta_fin}) and (\ref{eq:theta_osi}) to represent the variations in physical properties. Then a quantitative analysis between the non-uniformity of $\Theta$, denoted by $A_\theta$ and given by Eq. (\ref{eq:A_theta}), and the deviations from Boussinesq results are performed, resulting in a series of power laws given by Eqs. (\ref{eq:pl_1})-(\ref{eq:pl_end}). 
Besides, when the variable properties are included, the modified BA is reliable in both regimes, as the results calculated from it are in good agreement with those from full equations. It is believed that the modified BA incorporating variable physical properties is a simple but efficient model to describe double-diffusive convection in near-critical gases.

From a practical point of view, it is suggested that the variable properties are an important factor which should be taken into consideration in future studies and industrial applications. However, the availability of precise physical properties is very limited, especially for mixtures. The approximation for governing equations of near-critical gases should be made carefully. Besides, this work demonstrates that classical dynamic phenomena may exhibit new features in the critical region, implying that there are many potential research subjects to be explored.

\begin{acknowledgments}
	Zhan-Chao Hu and Xin-Rong Zhang gratefully acknowledge the support of National Natural Science Foundation of China (NO.51776002). Zhan-Chao Hu gratefully acknowledge financial support from China Scholarship Council (NO. 201706010268).
\end{acknowledgments}

\appendix
\section{Modeling of physical properties}\label{sec:phy_properties}
Near a critical point, a fluid or a fluid mixture exhibits large fluctuations of the order parameter associated with the critical-point phase transition. The range of the fluctuations can be characterized by a correlation length $\xi$. When the critical temperature $T_c$ is approached in the one-phase region at the critical density $\rho=\rho_c$, the correlation length diverges as \cite{Sengers1985}
\begin{equation}
\xi=\xi_0\varepsilon^{-0.63},
\end{equation}
where $\varepsilon = (T-T_c)/T_c$ is the reduced temperature, and $\xi_0$ is the amplitude (of the order of a molecular size). The divergence of correlation length posts a threat to the validity of Navier-Stokes equations. As noted by Zappoli \cite{ZAPPOLI2003713}, the limit of validity of the Navier-Stokes model is reached when the smallest macroscopic length introduced by the description is equal or smaller than the correlation length. The limit is estimated to be $\varepsilon= 10^{-6}$ for pure $\mathrm{CO_2}$. Since the molecular sizes of $\mathrm{CO_2}$ and $\mathrm{C_2H_6}$ are similar, a dynamic limit of $\varepsilon=10^{-6}$ is assumed in this study.

The critical fluctuations have strong effects on the thermodynamic properties of pure fluids and fluid mixtures. Asymptotically close to a critical point, the thermodynamic properties of pure fluids and fluid mixtures exhibit scaling laws with universal critical exponents. However, the scaling laws are valid in a limited range close to the critical point. A crossover model, connecting this and classical behavior far away from the critical point, has thus  been developed to represent the properties in the critical region \cite{Abbaci1991,Jin1993}. For a binary fluid, its  thermodynamic properties exhibit a crossover from one-component-like behavior to mixture-like asymptotic critical behavior. However, the mixture-like asymptotic scaling laws are not valid only until extremely close to a critical point ($\varepsilon\le10^{-8}-10^{-12}$) \cite{KISELEV199951}, which will not be encountered in this study due to the dynamic limit introduced earlier.

The divergence of correlation length also has profound effects on transport properties ($\eta$, $\lambda$, $D$). Far away from the critical point, transport properties are non-singular and generally slowly varying functions of temperature and density. Near the critical point, the divergence of correction length leads to a critical enhancement of transport properties. The asymptotic critical region, where simple power laws hold for transport properties, are much smaller than the region with the critical enhancement. Crossover models based on dynamic renormalization-group theory or mode-couping theory have been employed to model transport properties \cite{Luettmer1996}.

In this study, two approaches are considered to obtain the physical properties: the crossover model and a database called REFPROP \cite{LEMMON-RP91}.
\begin{table}
	\caption{\label{tab:prop} References for part of thermodynamic and transport properties.}
	\begin{ruledtabular}
		\begin{tabular}{cl}
			Property & References \\
			\hline
			$c_p$         & \cite{Luettmer1996} \\
			$c_v$         & \citet{Jin1993}\\
			$\alpha$    & \citet{Abbaci1991}\\
			$\eta$        & \citet{Luettmer1996}\\
			$\lambda$& \citet{Luettmer1996}\\
			$D$             & \citet{Luettmer1996}\\
		\end{tabular}
	\end{ruledtabular}
\end{table}

For the mixture considered in this study, available physical properties from crossover model are summarized in Table \ref{tab:prop}. In these references, the models have been tested against experimental data, and the properties are given as functions of $\varepsilon$ at $c=c_r$ and $\rho=\rho_c$. $\beta$ can be determined through the relation $\beta=\sqrt{\rho\alpha(c_p-c_v)/T}$. Besides, the critical parameters, $T_c$ and $\rho_c$, can be calculated from the relations developed by \citet{Sengers2007}. However, the data provided by the references is not sufficient. On the one hand, $\kappa$ is not available. On the other hand, only physical properties under $\rho=\rho_c$ and $c=c_r$ are available, while actually they are functions of $T$, $\rho$ and $c$. This is why REFPROP is considered. 

\begin{figure}
	\centering
	\includegraphics[width=12cm]{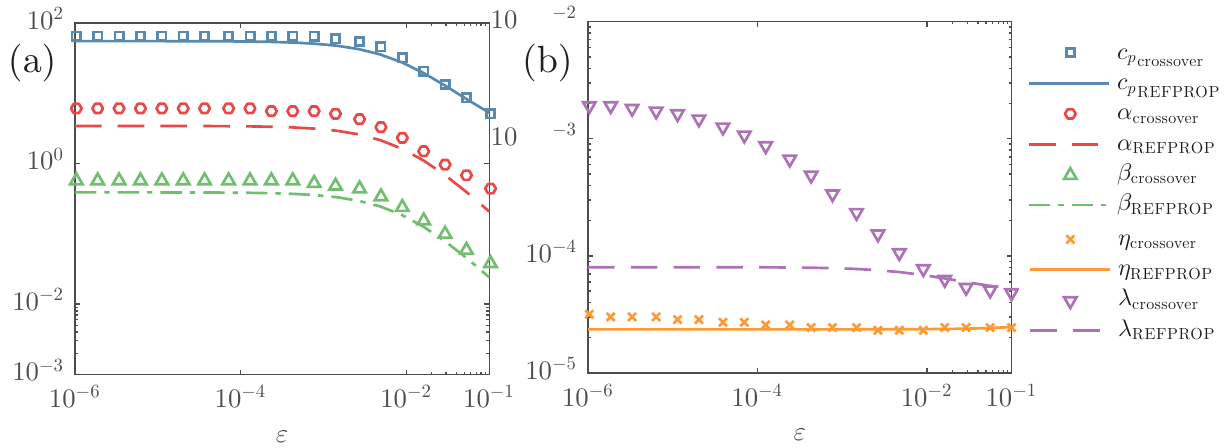}
	\caption{(Log-log plot) Several (a) thermodynamic and (b) transport properties of $\mathrm{CO_2}$-$\mathrm{C_2H_6}$ mixture ($c_r = 0.67$) along the critical isochore $\rho=\rho_c$, plotted against $\varepsilon$ from two sources: crossover model (see Table \ref{tab:prop}, note that $\beta=\sqrt{\rho\alpha(c_p-c_v)/T}$ ) and REFPROP \cite{LEMMON-RP91}. The units of these properties are $ c_p$: [$\rm{kJ/(kg\cdot K)}$], $\alpha$: [$\rm{MPa^{-1}}$], $\beta$: [$\rm{K^{-1}}$], $\eta$: [$\rm{Pa\cdot s} $], $\lambda$: [$\rm{kW/(m\cdot K)}$].}
	\label{fig:comp_refprop_crossover}
\end{figure}

For the $\mathrm{CO_2}$-$\mathrm{C_2H_6}$ mixture, all thermodynamic and transport properties except for $D$ can be calculated approximately by REFPROP through specifying $T$, $\rho$ and $c$. Figure \ref{fig:comp_refprop_crossover} shows and compares physical properties obtained from these two approaches. It is demonstrated that REFPROP can give good approximations for thermodynamic properties, where the behavior as the critical point is approached is the same: first experiencing a one-component-like strong divergence, and then remaining constant in a large crossover region \cite{KISELEV199951}.
The mixture-like asymptotic critical region is not shown because it is extremely close to the critical point. The critical region involved in this study can still be modeled by a classical equation of state (EOS) and a mixing law. The quantitative deviations between these two approaches are reasonable since different EOSs and mixing laws are embedded. However, for transport properties, since critical enhancement is not included in their modeling, REFPROP is reliable when the critical point is not closely approached.

\begin{figure}
	\centering
	\includegraphics[width=12cm]{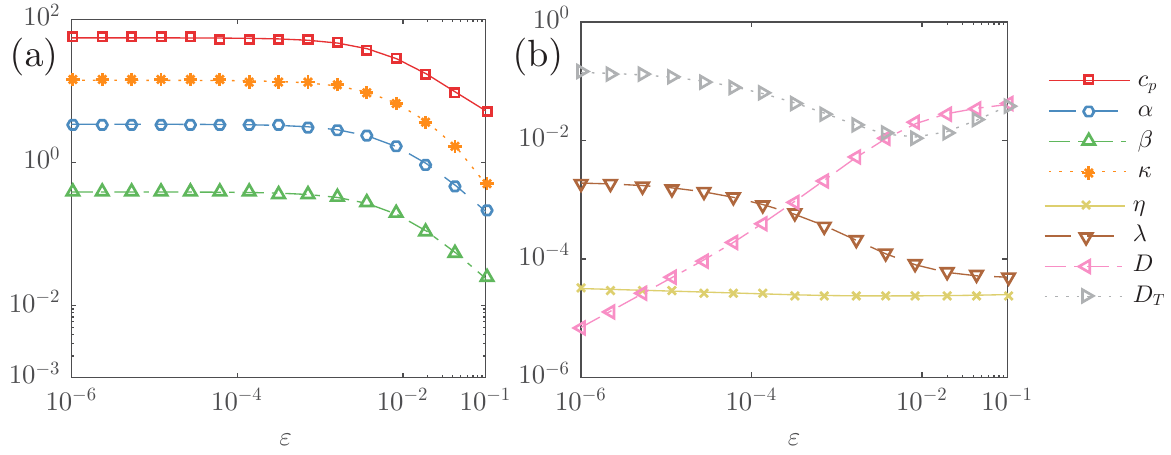}
	\caption{(Log-log plot)  (a) Thermodynamic and (b) transport properties of $\mathrm{CO_2}$-$\mathrm{C_2H_6}$ mixture ($c_r = 0.67$) along the critical isochore $\rho=\rho_r$, plotted against $\varepsilon$. $D_T=\lambda/(\rho c_p)$ is the thermal diffusivity. The units of these properties are $ c_p$: [$\rm{kJ/(kg\cdot K)}$], $\alpha$: [$\rm{MPa^{-1}}$], $\beta$: [$\rm{K^{-1}}$], $\kappa$: [$-$], $\eta$: [$\rm{Pa\cdot s} $], $\lambda$: [$\rm{kW/(m\cdot K)}$], $D$: [$\rm{mm^2/s}$], $D_T$: [$\rm{mm^2/s}$].}
	\label{fig:prop}
\end{figure}

Based on above discussions, the thermodynamic properties and the critical parameters are obtained from REFPROP. The transport properties, assumed as functions of $\varepsilon$, are determined through interpolating from data calculated by the crossover model \cite{Luettmer1996}. Note that since $T_c$ is a function of $c$, $\varepsilon$ is also a function of $c$. That is to say, only the dependence of density is not included in the modeling of transport properties. 
Figure \ref{fig:prop} plots all physical properties against $\varepsilon$ at $\rho=\rho_c$. Since the differences in rates of diffusion of heat and mass are crucial in DDC, the thermal diffusivity, defined as $D_T=\lambda/(\rho c_p)$, is also calculated and shown in Fig. \ref{fig:prop}(b). For transport properties, it is striking to find that $D$ tends to zero quickly as the critical point is approached, while $D_T$ develops a plateau. In addition, comparing to $\lambda$ and $D$, the variation of $\eta$ is small (less than one order).

\section{Several reduced cases}\label{appB}
\subsection{Case \RNum{1}: FC}
In FC, all physical properties are treated as constant values. As a result, Eq. (\ref{eq:LSA_var}) is simplified into 
\begin{eqnarray}\label{eq:case1}
\left( {{{\mathcal{D}}^2} - \frac{4}{3}{k^2}} \right)\hat u + \frac{1}{3}ik{\mathcal{D}}\hat w - ik\hat p &=& {\rm{ }}\sigma \left( {\frac{1}{{\rm Pr }}\bar \rho \hat u} \right), \nonumber \\
\frac{1}{3}ik{\mathcal{D}}\hat u +\left( {\frac{4}{3}{{\mathcal{D}}^2} - {k^2}} \right)\hat w + {\rm Ra_T} \hat T - {{\rm Ra_S}}\hat c - ({\mathcal{D}} + \chi )\hat p &=& \sigma \left( {\frac{1}{{\rm Pr }}\bar \rho \hat w} \right), \nonumber\\
\qquad\qquad \left[ {1 + \frac{{{\rm Ra_{ad}}}}{{\rm Ra_T}}\left(1-\frac{1 }{{\bar \rho }}\right)} \right]\hat w + \frac{1 }{{\bar \rho }}({{\mathcal{D}}^{\rm{2}}} - {k^2})\hat T &=& \sigma \left( {\hat T - \frac{1}{{{{\bar \rho }^2}}}\frac{{{ \Pi}{\rm Ra_{ad}}}}{{\rm Ra_T}}\hat p} \right),\\
\hat w + \tau ({{\mathcal{D}}^{\rm{2}}} - {k^2})\hat c + \frac{\tau }{{\bar \rho }}{\bar \rho_z}{\mathcal{D}}\hat c &=& \sigma \hat c,\nonumber\\
- ik\bar \rho \hat u -(\bar \rho_z  + \bar \rho {\mathcal{D}} ) \hat w&=& \sigma \left[ {{ \Pi}(\chi  \hat p - {\rm Ra_T}\hat T + {{\rm Ra_S}}\hat c)}\right]\nonumber.\qquad
\end{eqnarray}
The boundary condition is given by Eq. (\ref{eq:b_full}).

\subsection{Case \RNum{2}: BC}
In BC, the BA is applied, and all physical properties are treated as constant values. We define a stream function
\[
\tilde u  = \frac{\partial\tilde \psi}{\partial z}, ~~ \tilde w  = - \frac{\partial\tilde \psi}{\partial x}, 
\]
along with the normal mode expansion,
\[
\tilde\psi = \hat\psi(z){\rm exp}(ikx+\sigma t).
\]
The equations for perturbations after normal mode expansion are given by
\begin{eqnarray}\label{eq:case2}
({\mathcal{D}^4} - 2{k^2}{\mathcal{D}^2} + {k^4})\hat \psi- ik{\rm Ra_T}\hat T + ik{{\rm Ra_S}} \hat c &=&  \sigma \left[ {\frac{1}{{ \rm Pr  }}({\mathcal{D}^2} - {k^2})\hat \psi}\right], \nonumber \\
-ik\hat \psi + ({{\mathcal{D}}^{\rm{2}}} - {k^2})\hat T &=& \sigma \hat T ,\\
-ik\hat \psi + \tau({{\mathcal{D}}^{\rm{2}}} - {k^2})\hat c &=& \sigma \hat c.\nonumber
\end{eqnarray}
In the derivation of Eq. (\ref{eq:case2}), the incompressible continuity equation and two momentum equations have been combined and simplified into one single equation of $w$. The corresponding boundary condition is 
\begin{equation}\label{eq:b_Bou}
\hat \psi = \mathcal{D}\hat \psi = \hat T = \hat c = 0,  \ \quad \mbox{on\ }\quad z = 0 ~\&~ 1.
\end{equation}

\subsection{Case \RNum{3}: BC'}
Based on BC, BC' introduces the base-state density into the equation. The continuity equation $u_x+w_z=0$ suggests $\bar \rho_z = 0$, which is employed in the derivation. This can be seen as a modified BA, given by
\begin{eqnarray}\label{eq:case3}
({\mathcal{D}^4} - 2{k^2}{\mathcal{D}^2} + {k^4})\hat \psi- ik{\rm Ra_T}\hat T + ik{{\rm Ra_S}} \hat c &=&  \sigma \left[ {\frac{\bar\rho}{{\rm Pr }}({\mathcal{D}^2} - {k^2})\hat \psi}\right],  \nonumber \\
-ik\left[ {1 + \frac{{{\rm Ra_{ad}}}}{{\rm Ra_T}}\left(1-\frac{1 }{{\bar \rho }}\right)} \right]\hat \psi + \frac{1 }{{\bar \rho }}({{\mathcal{D}}^{\rm{2}}} - {k^2})\hat T &=& \sigma \hat T,\\
-ik\hat \psi + \tau({{\mathcal{D}}^{\rm{2}}} - {k^2})\hat c &=& \sigma \hat c. \nonumber
\end{eqnarray}
The boundary condition is given by Eq. (\ref{eq:b_Bou}).

\subsection{Case \RNum{5}: FV'}
Based on BC', BV' takes the variable physical properties into consideration, which leads to the following equations:
\begin{eqnarray}\label{eq:case5}
\eta ({\mathcal{D}^4} - 2{k^2}{\mathcal{D}^2} + {k^4})\hat \psi + 2{\eta _z}({\mathcal{D}^3}- {k^2}\mathcal{D})\hat \psi  && \nonumber \\
+{\eta _{zz}}({\mathcal{D}^2} + {k^2})\hat \psi - {ik}{\rm Ra_T}\beta \hat T + i{k}{{\rm Ra_S}}\kappa \hat c &=& \sigma \left[ {\frac{\bar\rho}{{ \rm Pr }}({\mathcal{D}^2} - {k^2})\hat \psi} \right], \nonumber\\
ik \left( {{{\bar T}_z} + \frac{{{\bar{{Ra}}_{ad}}}}{{\rm Ra_T}}\frac{\beta }{{{\bar \rho c_p}}}} \right)\hat \psi + \frac{\lambda }{{{\bar \rho c_p}}}({{\mathcal{D}}^{\rm{2}}} - {k^2})\hat T + \frac{{{\lambda _z}}}{{{\bar \rho c_p}}}{\mathcal{D}}\hat T &=& \sigma \hat T ,\\
ik {{\bar c}_z}\hat \psi + \tau D({{\mathcal{D}}^{\rm{2}}} - {k^2})\hat c + \frac{\tau }{{\bar \rho }}{(\bar \rho D)_z}{\mathcal{D}}\hat c &=& \sigma \hat c. \nonumber
\end{eqnarray}
The boundary condition is given by Eq. (\ref{eq:b_Bou}).

\bibliography{manuscript.bib}

\end{document}